\documentclass[aoas]{imsart}

\RequirePackage{amsthm,amsmath,amsfonts,amssymb}
\RequirePackage[authoryear]{natbib}
\RequirePackage[colorlinks,citecolor=blue,urlcolor=blue]{hyperref}
\RequirePackage{graphicx}
\usepackage{subfigure} 
\usepackage{fix-cm}
\usepackage{siunitx}
\usepackage{adjustbox}

\startlocaldefs
\def\V{\mbox{vec}}
\def\Vh{\mbox{vech}}
\def\ins{\small\mbox{ins}}
\def\bag{\small\mbox{bag}}

\def\sub{\small\mbox{sub}}

\def\MSE{\mbox{MSE}}
\def\test{\small\mbox{test}}
\newtheorem{theorem}{Theorem}
\def\diag{\mbox{diag}}
\def\cov{\mbox{cov}}

\endlocaldefs

\def\V{\mbox{vec}}

\def\mR{\mathbb{R}}

\def\cov{\mbox{cov}}

\def\diag{\mbox{diag}}

\usepackage{changes}
\usepackage{comment}
\usepackage{lineno}
\usepackage[normalem]{ulem}
\useunder{\uline}{\ul}{}

\usepackage{ulem}

\begin{document}

\begin{frontmatter}
\title{Detecting Breast Carcinoma Metastasis on Whole-Slide Images by Partially Subsampled Multiple Instance Learning\thanksref{T1}}
\runtitle{Partially Subsampled Multiple Instance Learning}
\thankstext{T1}{Xuetong Li is the corresponding author.}

\begin{aug}
\author[A]{\fnms{Baichen}~\snm{Yu}\ead[label=e1]{baichen.yu@stu.pku.edu.cn}},
\author[B]{\fnms{Xuetong}~\snm{Li}\ead[label=e2]{xtong\_li@xjtu.edu.cn}},
\author[C]{\fnms{Jing}~\snm{Zhou}\ead[label=e3]{zhoujing\_89@126.com}},
\and
\author[A]{\fnms{Hansheng}~\snm{Wang}\ead[label=e4]{hansheng@gsm.pku.edu.cn}}
\address[A]{Guanghua School of Management, Peking University, Beijing, China\printead[presep={,\ }]{e1,e4}}
\address[B]{School of Mathematics and Statistics, Xi'an Jiaotong University, Xi'an, China\printead[presep={,\ }]{e2}}
\address[C]{Center for Applied Statistics, School of Statistics, Renmin University of China, Beijing, China\printead[presep={,\ }]{e3}}
\end{aug}

\begin{abstract} 
Breast cancer is the most prevalent cancer in women worldwide. Histopathology image analysis serves as the gold standard for cancer diagnosis. In this regard, whole-slide imaging (WSI), a revolutionary technology in digital pathology, allows for ultrahigh-resolution tissue analysis. Despite its promise, WSI analysis faces significant computational challenges due to its massive data size and tissue heterogeneity. To address this issue, we present a Gaussian mixture based multiple instance learning (MIL) framework for WSI analysis with partially subsampled instances. Our approach models a WSI as a bag of instances (i.e., randomly cropped sub-images), leveraging a bag-based maximum likelihood estimator (BMLE) to predict metastases. Furthermore, we introduce a subsampling-based maximum likelihood estimator (SMLE) to refine predictions by selectively labeling a subset of instances. Extensive evaluations of the breast carcinoma metastasis prediction demonstrate that BMLE surpasses state-of-the-art methods, while the SMLE further improves the prediction accuracy at both bag and instance levels. We find that our method is fairly robust against various plausible model mis-specifications. Theoretical analyses and simulation studies validate the performance and robustness of our methods. 
\end{abstract}


\end{frontmatter}

\section{Introduction}\label{sec:intro}

Breast cancer is one of the most commonly diagnosed cancers in women globally.
As reported by \cite{Bray2024GlobalCS}, the year of 2022 has witnessed around 2.3 million novel cases and 666,000 fatalities resulting from this disease.
This high mortality rate underscores the urgency and importance of improving early detection and diagnosis.
In clinical practice, pathological needle biopsy is regarded as the gold standard for diagnosis \citep{waks2019breast}. Specifically, breast carcinoma is routinely identified by well-trained pathologists through checking of tissue slide stained with hematoxylin and eosin (H\&E) under high-power microscopy \citep{wang2019weakly}. Therefore, it is of great interest to automatically detect breast carcinoma in H\&E tissues. To this end, various disease pattern detection methods are needed.

Disease pattern detection refers to the identification of specific patterns or characteristics of disease development by analyzing various data and signals \citep{rangayyan2024biomedical}. 
In this regard, different types of disease pattern detection methods have been developed.
Particularly, various modern medical imaging techniques have been found useful \citep{mohammed2021radiohead,houssein2021deep}.
For example, \cite{patz2014overdiagnosis} used a convolution model for low-dose computed tomography (CT) and chest X-ray to explore the effect of various screening scenarios on cancer overdiagnosis.
\cite{liu2024robust} proposed a robust high-dimensional regression with coefficient thresholding for detecting disorders with functional magnetic resonance imaging (fMRI) data. 
A novel statistical disease-mapping framework was developed by \cite{liu2021statistical} to address the phenotypic and genetic heterogeneity across subjects and subpopulations. 

However, the existing techniques for disease pattern detection are mainly used to detect abnormal anatomy and physiology of the body \citep{withers2021x}.
They cannot observe details at the cellular level and thus cannot serve as the gold standard for cancer diagnosis.
To solve the problem, a new type of imaging technique, known as whole-slide image (WSI), has been developed and becomes increasingly popular for pathological diagnosis \citep{ghaznavi2013digital}.
Compared with traditional imaging techniques, WSI converts pathological slices containing tissue samples into digital images with ultrahigh resolution \citep{bejnordi2017diagnostic}.
The most distinct feature of WSI is its capacity to produce images with remarkably high resolution, often exceeding billions of pixels on one single slide \citep{zhu2017wsisa}.
This capability enables a meticulous examination of cellular architecture and tissue morphology, which is essential for precision in medical diagnostics.

To deal with the WSI data, many practitioners advocate the idea of multiple instance learning \citep[MIL,][]{hou2016patch}. 
Specifically, we treat each WSI sample as a bag, and label it by metastasis cell presence \citep{wang2022label}.
Next, assume that each bag contains multiple randomly cropped sub-images (i.e., instances), which are labeled by metastasis cell presence.
Here, the goal is to train a model using instance features and bag-level responses for accurate metastasis prediction.
Many methods have been developed for detecting breast cancer metastasis under an MIL framework.
\cite{carbonneau2018multiple} classified those methods into two groups. 
The first group contains methods which rely on fully observed instance-level labels.
As reported by \cite{bejnordi2017diagnostic},
those methods could be highly accurate in predicting at the instance level, but suffer from high labeling costs.
{The second group focuses on bag-level labels only, resulting in highly accurate bag-level predictions \cite{chen2024towards}. 
For example, \cite{campanella2019clinical} proposed a deep learning framework that combines convolutional neural networks with recurrent neural networks under an MIL approach.
\cite{lee2022derivation} introduced tumor-microenvironment-associated context learning using graph deep learning, which preserves the spatial relationships of each local feature. 
However, their instance-level prediction performance is often considerably worse. 
Meanwhile, detecting tumors at the instance level is of great clinical importance for patient diagnosis and monitoring \citep{ibrahim2020artificial}. 
Recently, the TransMIL method of \cite{shao2021transmil} encoded the spatial information of WSIs by the attention mechanism. They focused more on bag-level classification by aggregating the attention score.
\cite{wu2024individualized} proposed a deep logistic regression model at the patch level with total variation regularization. Then, how to achieve highly accurate prediction results at both bag and instance levels simultaneously becomes a problem of great importance.

To solve the problem, we develop here a Gaussian mixture based MIL framework. 
Specifically, we represent each instance by a feature vector, which is extracted by a pre-trained deep learning model. 
We next impose on those feature vectors a multivariate Gaussian mixture model (GMM). 
Similar methods utilizing Gaussian mixture models to describe instance-level features were developed by \cite{song2024morphological}, which takes a two-step procedure. In the first step, they convert a large amount of instance-level feature vectors into one single bag-level feature vector in a fully unsupervised way. In the second step, a downstream classification model based on this bag-level feature vector and bag-level labels is trained. In contrast, our method is a one-step and end-to-end learning method, which learns the probabilistic relationship between the bag-level label and a large amount of instance-level features directly in a weakly or supervised way. As a consequence, we develop a novel maximum likelihood estimator and refer to it as a bag-based maximum likelihood estimator (BMLE).}  
The BMLE can be computed by an appropriately designed EM algorithm.
The key merit of the BMLE is that only bag-level labels are needed \citep{li2018thoracic}. 
However, the price paid for this convenience is that the statistical efficiency of the BMLE could be sacrificed to a large extent.
To make better use of the rich information contained in all instances, we further propose a subsampling-based maximum likelihood estimator (SMLE).
The SMLE combines the information from both the bag-level labels and the partially labeled instances \citep{deng2024optimal}, which are selected by a carefully designed subsampling strategy. 
Asymptotic theories are developed for the theoretical understanding of various estimators (i.e., the IMLE, BMLE, and SMLE).
Simulation studies are conducted to demonstrate the finite sample performance of the proposed methods. The robustness of our methods against various plausible model assumption violations is evaluated both theoretically and numerically. 
The real data analysis on the CAMELYON16 dataset suggests that the SMLE performs better than the state-of-the-art (SOTA) result at both bag and instance levels.

The rest of the article is organized as follows. In Section \ref{sec:datades}, we provide an overview and the applied motivation of the CAMELYON16 dataset for the analysis. The main methodology is developed in Section \ref{sec:method}. 
Section \ref{sec:simu} presents the simulation studies to demonstrate finite sample performance of the proposed estimators. 
Section \ref{sec:robust} studies the robustness of our methods against the plausible violation of the model assumptions both theoretically and numerically.
Section \ref{sec:realdata} presents the real data analysis on the CAMELYON16 dataset.
Finally, the article is concluded with a brief discussion by Section \ref{sec:conclude} and a short significance statement by Section \ref{sec:sign}. All technical details and additional numerical studies are provided in the Appendices.

\section{The CAMELYON16 Dataset}\label{sec:datades}

In breast cancer treatment, detecting the spread of carcinoma to other organs and tissues is critically important. The sentinel lymph node acts as the ``sentinel'' for breast carcinoma, located in the axillary region and serving as the initial site for lymphatic drainage from the breast \citep{krag1998sentinel}. It serves as one of the most significant prognostic factors in breast carcinoma \citep{kawada2011significance}. Prognosis is poorer when carcinoma has metastasized to the lymph nodes. Therefore, the automatic detection of whether metastasis exists in the sentinel lymph nodes becomes a problem of great importance. Specifically, the metastasis detection should be done at both bag and instance levels. Bag-level prediction identifies the patient with breast carcinoma metastasis as well as its severity. Meanwhile, detecting metastases at the instance level helps pathological grading \citep{elmore2015diagnostic}, and this is essential for tracking disease progression, evaluating treatment efficacy, and detecting early recurrence \citep{ibrahim2020artificial}. Consequently, it is of clinical importance to predict metastases accurately at both bag and instance levels simultaneously.

To address this issue, \cite{bejnordi2017diagnostic} invested substantial effort in constructing the CAMELYON16 dataset with ultrahigh resolution. This is an important dataset for breast carcinoma metastasis detection on lymph nodes.
It is publicly available at \url{https://camelyon16.grand-challenge.org/}. 
More specifically, the CAMELYON16 dataset comprises a total of 399 whole-slide images (WSIs) of digitally scanned sentinel lymph node biopsy sections with ultrahigh resolution. Those WSIs were retrospectively sampled from the 399 patients, who underwent surgery for breast carcinoma at two hospitals in the Netherlands. They are, respectively, Radboud University Medical Center and University Medical Center Utrecht.
The dataset has already been partitioned into a training dataset and a testing dataset. The training dataset includes 270 WSIs, while the test dataset contains 129 WSIs. Unfortunately, some of the images are incorrectly labeled, unnecessarily duplicated, or suffer from incomplete annotations \citep{bejnordi2017diagnostic}.  
Those problematic images are then excluded from subsequent studies. 
This leads to a dataset with a total of $247$ images for training and $128$ images for testing. 

For this CAMELYON16 dataset, each WSI sample is either positive for including metastatic tumor cells or negative for not. Approximately 35.8\% of the training WSI slides and 38.0\% of the test slides are positive. Of the positive WSI samples, about 76.7\% contain infiltrating ductal carcinoma, and the remainder represent other histotypes. Moreover, WSI samples with both macrometastases (tumor cell cluster diameter $\geq$ \SI{2}{\milli\metre}) and micrometastases (tumor cell cluster diameter ranging from \SI{0.2}{\milli\metre} to \SI{2}{\milli\metre}) are included. Among positive training slides, 55.5\% exhibit macrometastases; among positive test slides, the percentage is 44.9\%. This study does not include WSI samples with isolated tumor cells (ITCs, single tumor cells or clusters with diameter $\leq$ \SI{0.2}{\milli\metre} or fewer than 200 cells) only, since the prognostic significance of ITCs remains unclear \citep{cucinella2024prognostic}. The widths of the images range from $45{,}056$ to $217{,}088$ pixels. The heights of the images range from $35{,}840$ to $221{,}696$ pixels. The average resolution level in terms of pixels is about $20$ billion per image. The average size of one single WSI on the hard drive is about $1.8$GB.

\begin{figure*}[!ht]
\centering
\includegraphics[scale=0.4]{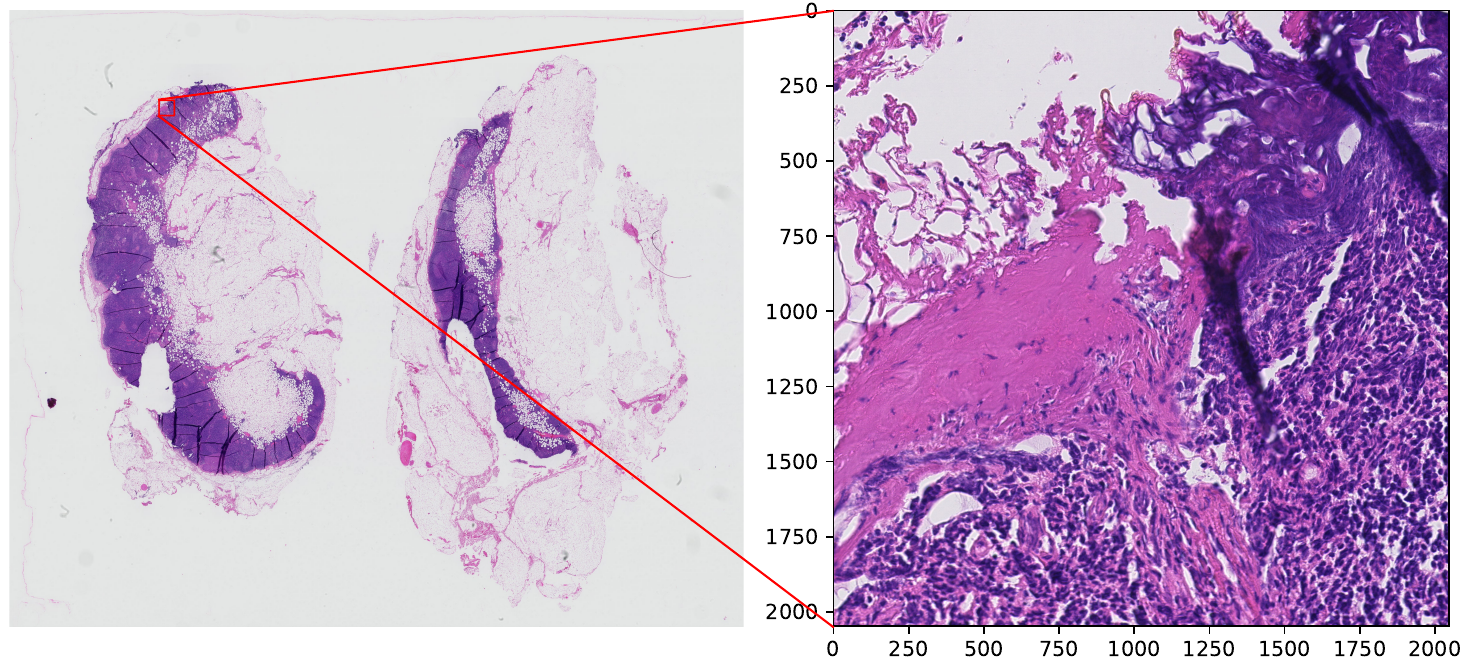}
\caption{The illustration of a WSI sample. The left panel is the origin $98{,}304\times 82{,}432$ WSI with a $2{,}048\times 2{,}048$ red bounding box. The right panel illustrates the zoomed-in sub-image of the bounded region in the left panel.}
\label{Fig:ImgShow}
\end{figure*}

Next, we should note that the statistical analysis of the CAMELYON16 dataset faces two serious challenges.
The first challenge is the massive size of the WSIs. As we mentioned above, each WSI is an ultrahigh-dimensional color image. For an intuitive understanding, one arbitrarily selected WSI sample is illustrated in Figure \ref{Fig:ImgShow}. 
The left panel of Figure \ref{Fig:ImgShow} is a WSI with detailed image information about metastasis tissue, normal tissue, stromal tissue, and others. 
The right panel in Figure \ref{Fig:ImgShow} represents a very small localized region of tissue with more image details.
This WSI sample consists of $98{,}304\times 82{,}432 = 8.10\times 10^{9}$ pixels, occupying approximately $1.1$ GB space on a hard drive. 
Therefore, it is practically an extremely difficult task if every single WSI is to be loaded into the CPU and/or GPU memory and processed as a whole \citep{stanisavljevic2018fast}. The massive size of the WSI data makes the statistical computation of the WSI data practically extremely difficult, unless novel statistical learning methods can be applied. 
The second challenge is the heterogeneity of tissue samples \citep{zhu2017wsisa,zhang2023image}. Different patients may have different disease progress, which leads to the heterogeneity in WSI tissue samples. 
In the CAMELYON16 dataset, the percentage of sub-images related to tumor on each positive WSI varies greatly with a mean of 6.61\% and a standard deviation of 24.85\%. 
Therefore, it is essential to develop a statistical method, which is robust against this heterogeneous issue.

\section{Methodology}\label{sec:method}
\subsection{Instance-based Maximum Likelihood Estimator}\label{sec:method-imle}

Consider a collection of $N$ independent bags (i.e., the WSI samples in our study) indexed by $1\leq i \leq N$. 
Let $Y_{i}\in \{0,1\}$ be the bag-level binary response of the $i$-th bag with $P(Y_{i}=1)=\alpha \in (0,1)$. 
Here $Y_{i} = 1$ if metastatic cells are detected in the $i$-th WSI sample, and $Y_{i} = 0$ otherwise \citep{wang2022label}.
Next, we assume a number of instances for each bag.
Here, an instance refers to a randomly cropped sub-image of the WSI sample with pre-specified sizes. 
For notation simplicity, we assume that the number of instances in each bag $i$ is the same and is given by $M$. 
Define $A_{im}\in\{0,1\}$ to be the binary label associated with the $m$-th instance in the $i$-th bag. 
Specifically, $A_{im} = 1$ if metastatic cells are detected in the $m$-th sub-image, and $A_{im} = 0$ otherwise.
If $Y_{i} = 0$, we should have $A_{im}=0$ for every $1\leq m\leq M$. 
{In contrast, if $Y_{i}=1$, we should have $A_{im}$s for different $1\leq m\leq M$ independently generated with $P(A_{im}=1|Y_{i}=1)=\pi$, where $\pi$ is an unknown parameter to be estimated.}

We next consider how to generate features. 
Given $A_{im}$, we generate a $p$-dimensional feature vector $X_{im}=(X_{imj})\in\mathbb{R}^{p}$ for each instance. 
This is a feature vector extracted from sub-images by some well developed deep learning methods. {Next, assume that the conditional distribution of $X_{im}$ with both $Y_{i}$ and $A_{i} = \{A_{im}:1\leq m\leq M\}$ given should depend on $A_{im}$ only.
Given $Y_{i}$ and $A_{im}$, we assume that $X_{im}$s are independently generated without spatial dependence. More specifically, we assume for $X_{im}$ with $A_{im}=k$ given a multivariate Gaussian distribution with mean $\mu_{k}=(\mu_{kj}:1\leq j\leq p)\in\mathbb{R}^{p}$ and covariance 
$\Sigma = (\sigma_{j_1j_2}:1\leq j_{1},j_{2}\leq p)\in\mathbb{R}^{p\times p}$. 
Define $\phi_{\mu,\Sigma}(x) = (2\pi)^{-p/2} |\Sigma|^{-1/2} \exp\big\{-(x-\mu)^{\top}\Sigma^{-1}(x-\mu)\big/2\big\}$ as the probability density function of the multivariate Gaussian distribution with mean $\mu\in\mathbb{R}^{p}$ and covariance 
$\Sigma=\big(\sigma_{j_{1}j_{2}}\big)\in\mathbb{R}^{p\times p}$. 
For an arbitrary square matrix $B=(b_{ij})\in\mathbb{R}^{p\times p}$, define a vectorization operator as $\V(B) = (b_{ij}:1\leq i, j\leq p)\in\mathbb{R}^{p^2}$. 
Moreover, if $B$ is a symmetric matrix, define a half-vectorization operator as $\Vh(B) = (b_{ij}:1\leq i\leq j\leq p)\in\mathbb{R}^{p(p+1)/2}$. Then, there should exist a unique constant matrix $D\in\mathbb{R}^{p^2\times p(p+1)/2}$ such that $\V(B)=D\Vh(B)$. Explicitly, we have $D = \sum_{i \geq j} \V (T_{ij})u_{ij}^{\top}$, where $u_{ij}\in\mathbb{R}^{p(p+1)/2}$ with the value $1$ in the position $(j-1)p + i - j(j-1)/2$ and $0$ elsewhere, and $T_{ij}\in\mathbb{R}^{p \times p}$ with $1$ in positions $(i,j)$ and $(j,i)$ and zeros elsewhere. See \cite{magnus2019matrix} for more technical details. 
For better modeling flexibility, we can modify our model slightly, so that the heterogeneous covariance structure can be used.
This modification causes little difficulty to our subsequent algorithm development; see Appendix A.4 for the details. 
However, it unfortunately leads to degraded predictive performance on the CAMELYON16 dataset. 
This is because a large number of extra parameters are inevitably involved.
Therefore, we should keep this homogeneity assumption of variances in the following sections for better predictive performance on the CAMELYON16 dataset.}

Note that $\alpha$ can be easily estimated by $\widehat{\alpha} = N^{-1}\sum_{i=1}^{N} Y_{i}$. 
Therefore, we should subsequently focus on the estimation of parameters $\pi$, $\mu_{1}$, $\mu_{0}$, and $\Sigma$. 
Write $\Omega = \Sigma^{-1}\in\mathbb{R}^{p\times p}$ as the precision matrix. 
Define $\Theta = \big(\pi, \mu_{1}^{\top},\mu_{0}^{\top}, \Vh(\Omega)^{\top}\big)^{\top}\in\mathbb{R}^{q}$, where $q = p^2/2+5p/2+1$ is the total number of parameters contained in $\Theta$. We start with the most ideal scenario, where the instance-level indicator $A_{im}$s are actually observed. 
Then, the instance-level log-likelihood can be analytically spelled out as
\begin{gather}
\mathcal{L}^{\ins}(\Theta) =  \sum_{i=1}^{N}\sum_{m=1}^{M}Y_{i}\Big\{A_{im}\log\pi + (1-A_{im}) \log(1-\pi)\Big\}\notag\\
+ \sum_{i=1}^{N} \sum_{m=1}^{M} \Big(1-A_{im}\Big)\log \phi_{\mu_{0},\Sigma}\Big(X_{im}\Big) + \sum_{i=1}^{N} \sum_{m=1}^{M} A_{im}\log \phi_{\mu_{1},\Sigma}\Big(X_{im}\Big).
\label{ins-log-lik}
\end{gather}
The instance-based maximum likelihood estimator (IMLE) can be then obtained by
\begin{gather*}
\widehat{\pi}^{\ins} = \Big(M\sum\limits_{i=1}^{N}Y_{i}\Big)^{-1}\sum\limits_{i=1}^{N} \sum\limits_{m=1}^{M} A_{im},\quad \widehat{\mu}_{1}^{\ins} = \Big(\sum\limits_{i=1}^{N} \sum\limits_{m=1}^{M}A_{im}\Big)^{-1}\Big(\sum\limits_{i=1}^{N} \sum\limits_{m=1}^{M} A_{im}X_{im}\Big),\\
\widehat{\mu}_{0}^{\ins} = \Big\{\sum\limits_{i=1}^{N} \sum\limits_{m=1}^{M}\big(1-A_{im}\big)\Big\}^{-1}\Big\{\sum\limits_{i=1}^{N} \sum\limits_{m=1}^{M} \big(1-A_{im}\big)X_{im}\Big\},\\
\widehat{\Sigma}^{\ins} = (NM)^{-1}\sum\limits_{i=1}^{N} \sum\limits_{m=1}^{M}\Big\{A_{im}\big(X_{im}-\widehat{\mu}_{1}^{\ins}\big)\big(X_{im}-\widehat{\mu}_{1}^{\ins}\big)^{\top} \\ 
+ \big(1-A_{im}\big)\big(X_{im}-\widehat{\mu}_{0}^{\ins}\big)\big(X_{im}-\widehat{\mu}_{0}^{\ins}\big)^{\top}\Big\}.
\end{gather*}
Define $\widehat{\Theta}^{\ins} = \big(\widehat{\pi}^{\ins}, \widehat{\mu}_{1}^{\ins\top},\widehat{\mu}_{0}^{\ins\top},  \Vh(\widehat{\Omega}^{\ins})^{\top}\big)^{\top}\in\mathbb{R}^{q}$ as the IMLE, where $\widehat{\Omega}^{\ins} = \big(\widehat{\Sigma}^{\ins}\big)^{-1}$. 
Then, $\widehat{\Theta}^{\ins}$ serves as an important benchmark for subsequent theoretical comparison.

Write $\mathcal{A} = \cup_{i=1}^{N} A_{i} = \{a_{im}: 1\leq i\leq N, 1\leq m\leq M\}$. Then, conditional on $\mathcal{A}$, we have $\widehat{\pi}^{\ins}$, $\widehat{\mu}_{1}^{\ins}$, $\widehat{\mu}_{0}^{\ins}$, and $\widehat{\Sigma}^{\ins}$ to be mutually independent with each other \citep{anderson1958introduction}. However, these quantities (i.e., $\widehat{\pi}^{\ins}$, $\widehat{\mu}_{1}^{\ins}$, $\widehat{\mu}_{0}^{\ins}$, and $\widehat{\Sigma}^{\ins}$ ) are all dependent on $\mathcal{A}$ marginally. Therefore, they are not strictly independent of each other in terms of finite sample distribution. Nevertheless, whether they are mutually independent of each other asymptotically is not immediately clear. 
{Define $\otimes$ to represent the Kronecker product.
For arbitrary matrices $A=(A_{i,j}) \in \mR^{m_1 \times n_1}$ and $B=(B_{k,l}) \in \mR^{m_2 \times n_2}$, their Kronecker product $C = A \otimes B \in \mR^{m_1n_1 \times m_2n_2}$ is defined entry-wise as $C_{(i-1)p + k, (j-1)q + l} = A_{i,j} \times B_{k,l}$.} We therefore develop the following theorem to describe the asymptotic behavior of $\widehat{\Theta}^{\ins}$.
\begin{theorem}
\label{thm-1}
We have $\sqrt{NM}\big(\widehat{\Theta}^{\ins}-\Theta\big)\rightarrow_{d} N\big\{0,(\Omega^{\ins})^{-1}\big\}$ as $\min(N,M)\rightarrow\infty$, where $\Omega^{\ins} = 
\diag\big\{\alpha\pi^{-1}(1-\pi)^{-1}, \alpha\pi\Omega, 
(1-\alpha\pi)\Omega, 
2^{-1}D^{\top} (\Sigma \otimes \Sigma) D \big\} \in \mathbb{R}^{q\times q}$, and ``$\rightarrow_{d}$" stands for converging in distribution. 
\end{theorem}
\noindent The detailed proof of Theorem \ref{thm-1} is provided in Appendix C.1. 
{Note that $\Sigma$ can be practically estimated by $\widehat{\Sigma}^{\ins}$, $\Omega$ by $\big(\widehat{\Sigma}^{\ins}\big)^{-1}$, $\alpha$ by $\widehat{\alpha}$, and $\pi$ by $\widehat{\pi}^{\ins}$. This leads to an estimated precision matrix $\widehat{\Omega}^{\ins} = \diag\big\{\widehat{\alpha}(\widehat{\pi}^{\ins})^{-1}(1-\widehat{\pi}^{\ins})^{-1}, \widehat{\alpha}\widehat{\pi}^{\ins}(\widehat{\Sigma}^{\ins})^{-1}, (1-\widehat{\alpha}\widehat{\pi}^{\ins})(\widehat{\Sigma}^{\ins})^{-1}, 2^{-1}D^{\top} (\widehat{\Sigma}^{\ins} \otimes \widehat{\Sigma}^{\ins}) D \big\} \in \mathbb{R}^{q\times q}$ for the asymptotic precision matrix $\Omega^{\ins}$. Therefore, by Slutsky's Theorem and the Continuous-Mapping Theorem \citep{van2000asymptotic}, Theorem \ref{thm-1} can be re-write as $(\widehat{\Omega}^{\ins})^{1/2}\sqrt{NM}\big(\widehat{\Theta}^{\ins}-\Theta\big)\rightarrow_{d} N(0,I_{q})$. We then obtain a studentized asymptotic distribution with a variance $I_{q}$ that does not depend on the parameters being estimated.} 
From Theorem \ref{thm-1}, we find that different components of $\widehat{\Theta}^{\ins}$ (i.e., $\widehat{\pi}^{\ins}$, $\widehat{\mu}_{1}^{\ins}$, $\widehat{\mu}_{0}^{\ins}$ and $\widehat{\Omega}^{\ins}$) are independent with each other asymptotically. 
Consider two special cases. The first special case is $\alpha=1$. In this case, all the bags are positive. 
Therefore, the problem reduces to a standard GMM with observed labels at the instance level. 
In this case, the asymptotic distribution of $\widehat{\Theta}^{\ins}$ has been given in \cite{anderson1958introduction}. 
The second special case is $\alpha=0$. In this case, all bags and thus the instances are negative. 
Therefore, neither $\pi$ nor $\mu_{1}$ can be estimated consistently. 
However, both the parameters $\mu_{0}$ and $\Sigma$ are not much affected.

\subsection{Bag-based Maximum Likelihood Estimator}\label{sec:method-bmle}

The IMLE discussed in Section \ref{sec:method-imle} is easy to compute and enjoys nice theoretical properties. 
Unfortunately, in real practice, the binary instance-level indicator $A_{im}$s are often too expensive to be fully observed. 
Instead, we often observe the bag-level response $Y_{i}$s only. 
Accordingly, we can only obtain the bag-level log-likelihood function as
\begin{gather}
\mathcal{L}^{\bag}(\Theta) = \sum\limits_{i=1}^{N} \Big(1-Y_{i}\Big)\sum\limits_{m=1}^{M} \log \phi_{\mu_{0},\Sigma}\Big(X_{im}\Big)\notag\\ + \sum\limits_{i=1}^{N} Y_{i} \sum\limits_{m=1}^{M} \log\bigg\{\pi\phi_{\mu_{1},\Sigma}\Big(X_{im}\Big) + \Big(1 - \pi\Big)\phi_{\mu_{0},\Sigma}\Big(X_{im}\Big)\bigg\}.
\label{log-like}
\end{gather}
Subsequently, a bag-based maximum likelihood estimator (BMLE) can be defined as $\widehat{\Theta}^{\bag} = \arg\max_{\Theta}\mathcal{L}^{\bag}(\Theta) = \big(\widehat{\pi}^{\bag}, \widehat{\mu}_{1}^{\bag\top},\widehat{\mu}_{0}^{\bag\top},\Vh(\widehat{\Omega}^{\bag})^{\top}\big)^{\top}\in\mathbb{R}^{q}$, where $\widehat{\Omega}^{\bag} = \big(\widehat{\Sigma}^{\bag}\big)^{-1}$ is the estimated precision matrix. {To numerically compute $\widehat{\Theta}^{\bag}$, an Expectation-Maximization (EM) algorithm can be developed with the details given in Appendix A.1.}
We next study the asymptotic properties of the BMLE $\widehat{\Theta}^{\bag}$ by the following theorem. 
\begin{theorem}\label{thm-2}
We have $\sqrt{NM}\big(\widehat{\Theta}^{\bag} - \Theta\big)\rightarrow_{d}N\big\{0,\big(\Omega^{\bag}\big)^{-1}\big\}$ as $\min(N, M)\rightarrow\infty$.
\end{theorem}
\noindent 
The analytical details of $\Omega^{\bag}$ and the detailed proof of Theorem \ref{thm-2} are provided in Appendix C.2. 
By Theorem 2, we know that the BMLE $\widehat{\Theta}^{\bag}$ is $\sqrt{NM}$-consistent and asymptotically normal. 
However, it has a different asymptotic covariance matrix as compared with that of the IMLE $\widehat{\Theta}^{\ins}$. Therefore, it is of great interest to compare their relative efficiency. Intuitively, we should expect that $\widehat{\Theta}^{\ins}$ to be statistically more efficient than $\widehat{\Theta}^{\bag}$, since $\widehat{\Theta}^{\ins}$ enjoys more information provided by the instance-level labels as compared with $\widehat{\Theta}^{\bag}$. 
However, the theoretical difference between $\widehat{\Theta}^{\ins}$ and $\widehat{\Theta}^{\bag}$ in terms of the asymptotic efficiency is not immediately clear. 
To gain some understanding, we can analyze their difference in terms of their asymptotic precision matrices (instead of the asymptotic variance matrices) due to their analytical simplicity. {More specifically, define this difference as $\Delta_{\bag}^{\ins} = \Omega^{\ins} - \Omega^{\bag}$  and $\widetilde{\pi}_{im} = P(A_{im}=1|X_{im},Y_{i})$ to be the posterior probability of $A_{im}$ given both $X_{im}$ and $Y_{i}$.} 
It can then be verified that $\Delta_{\bag}^{\ins} = E\big\{\widetilde{\pi}_{im}(1-\widetilde{\pi}_{im})\delta_{1}\delta_{1}^{\top}\big\}$, where 
$\delta_{1} = \big\{\pi^{-1}(1-\pi)^{-1}, (X_{im}-\mu_{1})^{\top} \Omega, -(X_{im}-\mu_{0})^{\top} \Omega, -2^{-1}D^\top \V((X_{im}-\mu_{1})(X_{im}-\mu_{1})^{\top} - (X_{im}-\mu_{0})(X_{im}-\mu_{0})^{\top})^{\top}\big\}^{\top} \in\mathbb{R}^{q}$. 
The verification details are given in Appendix C.3. 
Note that $\Delta_{\bag}^{\ins}$ is a positive semi-definite matrix. 
Therefore, as one might expect, $\widehat{\Theta}^{\ins}$ is statistically more efficient than $\widehat{\Theta}^{\bag}$. However, it seems that the difference should vanish if $\widetilde{\pi}_{im}\big(1-\widetilde{\pi}_{im}\big)\rightarrow 0$. Interestingly, this happens to be the case, where positive and negative instances can be differentiated from each other easily. This is because $\widetilde{\pi}_{im}\big(1-\widetilde{\pi}_{im}\big)\rightarrow 0$ implies that either $\widetilde{\pi}_{im}\rightarrow 0$ or $\widetilde{\pi}_{im}\rightarrow 1$. This implies that the positive and negative cases can be differentiated from each other easily.

\subsection{Subsampling-based Maximum Likelihood Estimator}\label{sec:method-smle}

The merit of the bag-level learning mentioned above is that only bag-level labels are needed for instance-level prediction \citep{li2018thoracic}. 
Note that the labeling cost needed for each bag is considerably lower than that of the instance-level labels \citep{liu2012key}. 
This is mainly due to the following reasons.
Firstly, to claim $Y_{i}=0$ for a given whole image, pathologists have to examine every $A_{im}$ and make sure that $A_{im}=0$ for every instance $1\le m \le M$.
This makes the effort to claim a bag-level label $Y_{i}=0$ and all instance-level labels $A_{im}=0$ for every instance $1\le m \le M$ nearly identical.
Nevertheless, the story changes dramatically if the pathologists wish to claim a positive case with $Y_{i}=1$.  
In this case, as long as one positive instance (i.e., a metastatic sub-region) is detected, the pathologists can stop the examination process immediately. 
This leads to a significant reduction in labeling cost, since the size of WSI is often huge and thus the total number of $A_{im}$s is often large.
This makes detecting the $A_{im}$s for every $1\le m \le M$ much harder than $Y_{i}$s for the case with $Y_{i}=1$. 
However, there is no free lunch. 
The price paid for this convenience is that the statistical efficiency of $\widehat{\Theta}^{\bag}$ could be sacrificed to some extent.
If the desired estimation accuracy is high, the total number of bags must be relatively large. 
This might be a practically appealing solution, if the bag sample size can be easily increased. 
Unfortunately, this is often not the case in real practice.
Instead, the bag samples are often very difficult or costly to obtain \citep{zhou2023ensemble}. 
As we mentioned in Section \ref{sec:datades}, in the CAMELYON16 dataset, each bag sample corresponds to a WSI sample, which corresponds to one patient who has undergone surgery treatment in one of the two hospitals in the Netherlands \citep{bejnordi2017diagnostic}. 
Since the number of patients cannot be easily increased, the bag sample size cannot be arbitrarily enlarged.

Since arbitrarily enlarging the bag sample size is often practically infeasible, how to make better use of the rich information contained in instances (without enlarging the bag sample size) becomes a promising direction.
To this end, we develop here a subsampling-based method. 
Note that the negative bags contain negative instances only. 
Comparatively, the positive instances are much more scarce and thus valuable. Therefore, we are motivated to provide more labels for those instances contained in positive bags with particular interest in positive instances. Specifically, define for the $m$-th instance in the $i$-th bag a binary subsampling indicator $S_{im}\in\{0,1\}$. Note that we are more interested in subsampling those positive instances, since a large number of negative instances have been provided by the negative bags already. Since we only subsample the instances from a positive bag, we should have $S_{im}=0$ for every $Y_{i}=0$.
Then, how to design the subsampling probability $P(S_{im}|X_{im},Y_{i}=1)$ for positive bags becomes the critical issue.

Ideally, the subsampling probability should be monotonically related to the positive instance probability $\pi_{im} = P(A_{im}=1|X_{im},Y_{i}=1)$. In other words, the more likely an instance is to be positive, the more possible it should be sampled. Under our model assumption, it can be easily verified that
\begin{align}
\pi_{im} = \frac{\pi\phi_{\mu_{1},\Sigma}(X_{im})}{\pi\phi_{\mu_{1},\Sigma}(X_{im}) + (1-\pi)\phi_{\mu_{0},\Sigma}(X_{im})} 
= \frac{\exp\big(\alpha_{0}+X_{im}^{\top}\beta\big)}{1 + \exp\big(\alpha_{0}+X_{im}^{\top}\beta\big)} , \label{posterior-prob}
\end{align}
where $\alpha_{0}=\big(\mu_{0}^{\top}\Omega\mu_{0} - \mu_{1}^{\top}\Omega\mu_{1}\big)\big/2 + \log\big\{(1-\pi)^{-1}\pi\big\}$ and $\beta = \Omega(\mu_{1}-\mu_{0})$. In other words, the instance-level positive probability $\pi_{im}$ is a monotonically increasing function in $X_{im}^{\top}\beta$. Therefore, we are motivated to design a subsampling strategy for $S_{im}$ as
\begin{align}
P\Big(S_{im}=1\Big|X_{im},Y_{i}=1\Big) = \frac{\exp\big(\alpha_{n}+X_{im}^{\top}\beta\big)}{1 + \exp\big(\alpha_{n}+X_{im}^{\top}\beta\big)} = \gamma_{im}.
\label{sampling-prob}
\end{align}
Compared with $\pi_{im}$, the only difference about $\gamma_{im}$ is that the model parameter $\alpha_{0}$ in \eqref{posterior-prob} is replaced by the hyperparameter $\alpha_{n}$ in \eqref{sampling-prob}. The former is initially determined by the population distribution about $X_{im}$ and therefore cannot be artificially modified. In contrast, the latter is a human-specified hyperparameter, which is used for controlling the overall subsampling fraction $\gamma = E\big(\sum_{i=1}^{N}\sum_{m=1}^{M}S_{im}\big|X_{im}, Y_{i}\big)\big(M\sum_{i=1}^{N}Y_{i}\big)^{-1} 
= \big(M\sum_{i=1}^{N}Y_{i}\big)^{-1}\big(\sum_{i=1}^{N}\sum_{m=1}^{M} Y_{i}\gamma_{im}\big)$. Ideally, the optimal value of $\alpha_n$ should be as large as possible, since a larger $\alpha_n$ leads to a larger portion of annotated samples and then better statistical efficiency.
Therefore, one should set $\alpha_n$ as large as possible in real practice, as long as one can afford the associated time and financial cost. 
To practically implement this strategy, the unknown parameter $\beta$ in \eqref{sampling-prob} needs to be estimated. 
Note that it can be easily estimated by $\widehat{\beta}=\widehat{\Omega}^{\bag}\big(\widehat{\mu}_{1}^{\bag}-\widehat{\mu}_{0}^{\bag}\big)$. 
That leads to the plug-in subsampling probability $\widehat{\gamma}_{im}= \exp\big(\alpha_{n}+X_{im}^{\top}\widehat{\beta}\big) \big/ \big\{1 + \exp(\alpha_{n}+X_{im}^{\top}\widehat{\beta})\big\}$. 
We then generate $S_{im}$s according to 
$P(S_{im}| X_{im},Y_{i}=1,\widehat\beta )=\widehat{\gamma}_{im}$. 
Given the instance-level labels obtained by the above subsampling strategy, we should be able to estimate the parameters of interest more accurately than the BMLE $\widehat{\Theta}^{\bag}$. 

Accordingly, we develop below an estimator based on the following likelihood function. 
The idea is to combine the information from both the bag-level labels and the partially labeled instances to form a more accurate estimator. The subsample-based log-likelihood function is given as
\begin{gather}
\mathcal{L}^{\sub}(\Theta) = \sum_{i=1}^{N} Y_{i} \sum_{m=1}^{M} \bigg[ \log \Big\{ \pi\phi_{\mu_{1},\Sigma}\big(X_{im}\big) + \big(1-\pi\big) \phi_{\mu_{0} \Sigma} \big(X_{im}\big) \Big\} + S_{im}\Big\{A_{im}\log\pi_{im} \notag\\ 
+ \big(1-A_{im}\big)\log\big(1-\pi_{im}\big)\Big\}\bigg] + 
\sum_{i=1}^{N} \Big(1-Y_{i}\Big) \sum_{m=1}^{M} \log\phi_{\mu_{0},\Sigma}\big(X_{im}\big).\label{log-like-2}
\end{gather}
Note that the log-likelihood function \eqref{log-like-2} contains the information from both the subsampled instances and the bag-level labels, which is analytically derived in Appendix A.2.
This leads to a subsample-based maximum likelihood estimator (SMLE) as $\widehat{\Theta}^{\sub} = \arg\max_{\Theta} \mathcal{L}^{\sub}(\Theta)$. To numerically compute $\widehat{\Theta}^{\sub}$, another EM algorithm can be developed with the details given in Appendix A.3.
Subsequently, the theoretical properties of $\widehat{\Theta}^{\sub}$ are given by the following theorem. 
\begin{theorem}\label{thm-4}
We have $\sqrt{NM}\big(\widehat{\Theta}^{\sub} - \Theta\big)\rightarrow_{d}N\big\{0,(\Omega^{\sub})^{-1}\big\}$ as $\min(N, M)\rightarrow\infty$.
\end{theorem}
\noindent
The analytical details of $\Omega^{\sub}$ and the detailed proof of Theorem \ref{thm-4} are provided in Appendix C.4. 
Theorem \ref{thm-4} leads to a number of interesting findings. First, we find that $\widehat{\Theta}^{\sub}$ remains $\sqrt{NM}$-consistent and asymptotically normal, even though those annotated instances included in \eqref{log-like-2} are not selected in a completely random manner. This is mainly because our subsampling indicator $S_{im}$ is a random variable depending on $X_{im}$s and $Y_{i}$s only. It is remarkable that, conditional on $X_{im}$s and $Y_{i}$s, $S_{im}$ and $A_{im}$ are mutually independent. Meanwhile, the only component related to the subsampling instances in \eqref{log-like-2} is those observed log-likelihood $\big\{A_{im}\log\pi_{im}+(1-A_{im})\log(1-\pi_{im})\big\}$ for every $(i,m)$ with $S_{im}=1$. As one can see, this is a conditional log-likelihood function with $A_{im}$ playing the role of response and $(X_{im}, Y_{i})$s playing the role of covariates. This makes our subsampling mechanism (as represented by $S_{im}$) missing at random, instead of nonignorable missing \citep{rubin1976inference,little2019statistical}. 
Second, from the proof in Appendix C.4, we find that the conclusion of Theorem \ref{thm-4} remains valid, as long as $\widehat\beta$ is consistent. 
Therefore, it is not necessary to use the full sample size to estimate $\beta$.
Instead, we can consider a reduced sample size for estimating $\widehat\beta$ so that the initial computational cost can be greatly reduced.
The resulting final estimator $\widehat\Theta^{\sub}$ should maintain the same asymptotic efficiency, which should be numerically validated in subsequent simulation studies. 
Last, we compare the relative efficiency of the SMLE $\widehat{\Theta}^{\sub}$ with that of the BMLE $\widehat{\Theta}^{\bag}$ obtained in Section \ref{sec:method-bmle}. 
To this end, define $\Delta_{\bag}^{\sub} = \Omega^{\sub}-\Omega^{\bag}$. 
Similar to Appendix C.3, it can be verified that $\Delta_{\bag}^{\sub} = E\big\{\gamma_{im}\widetilde{\pi}_{im}(1-\widetilde{\pi}_{im})\delta_{1} \delta_{1}^{\top} \big\}$. 
This is a positive semi-definite matrix, which implies that the SMLE $\widehat{\Theta}^{\sub}$ is statistically more efficient than the BMLE $\widehat{\Theta}^{\bag}$ uniformly. 
This is not surprising, since $\widehat{\Theta}^{\bag}$ utilizes the information provided by bag-level labels only. In contrast, $\widehat{\Theta}^{\sub}$ also benefits from the sub-sampled instances. The relative difference in terms of $\Delta_{\bag}^{\sub}$ vanishes as the subsampling fraction $\gamma$ goes to $0$. In this case, the SMLE becomes the BMLE. On the other hand, if the subsampling fraction $\gamma$ approaches $1$, then the SMLE $\widehat{\Theta}^{\sub}$ becomes the IMLE. This interesting theoretical finding is to be numerically verified in {\sc Study 4} of the subsequent section.

\begin{figure}[!ht]
\centering
\includegraphics[scale=0.3]{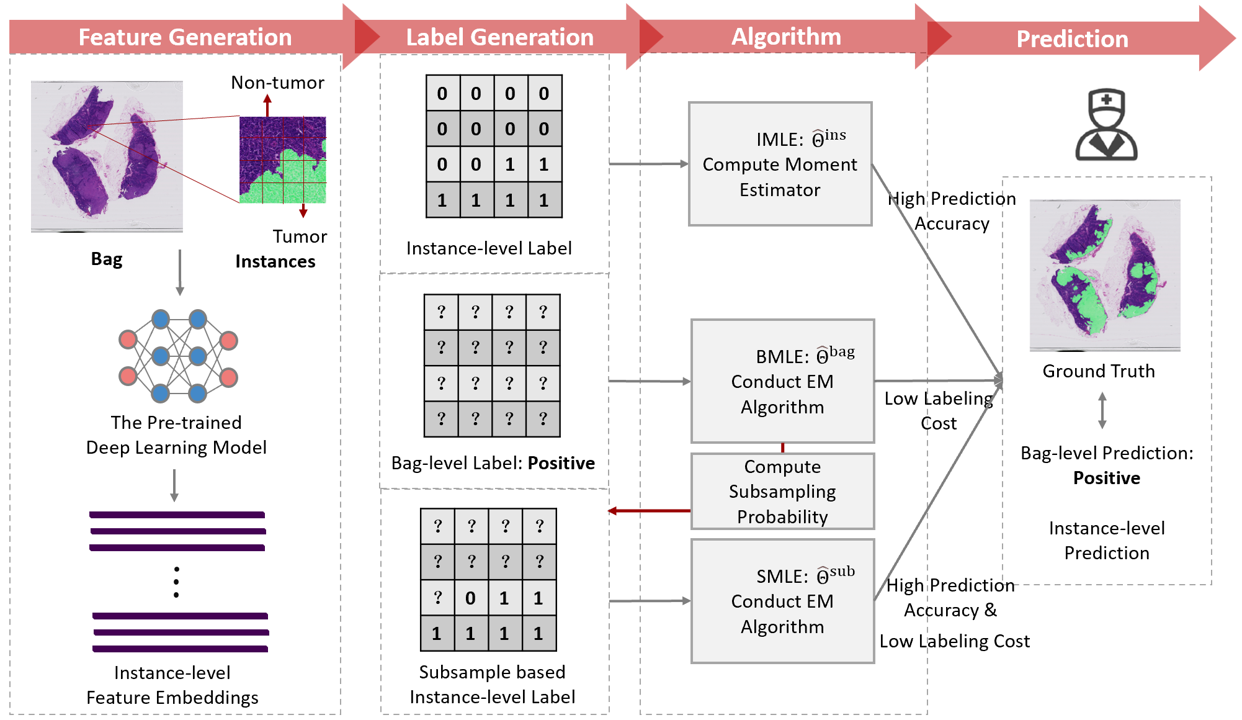}
\caption{The illustration of the pipeline. }
\label{Fig:pipeline}
\end{figure}

Finally, to comprehensively illustrate the application of our methods, we present a detailed pipeline in Figure \ref{Fig:pipeline}. This pipeline provides a clear and sequential overview of the entire process. 
First, each WSI is considered as a bag containing multiple instances. 
Each instance corresponds to a randomly cropped sub-image from the WSI. 
These instances are then converted into feature vectors extracted by a pre-trained deep learning model.
Second, we generate the corresponding instance-level and bag-level labels.
Specifically, a bag-level label can be defined based on whether metastasis cells exist in the WSI.
An instance-level label can be defined based on whether metastasis cells exist in the sub-image.
Third, we implement our three proposed algorithms depending on different levels of labels. Lastly, we obtain the prediction results.

\section{Simulation Studies}\label{sec:simu}

To demonstrate the finite sample performance of the various estimating methods, we present here a number of simulation studies. Since our research is motivated by this CAMELYON16 dataset which is also our intended application, we are therefore motivated to design our simulation study in a way most similar to the CAMELYON16 dataset. Specifically, the bag-level positive probability $\alpha$ and instance-level positive probability $\pi$ are set to be their empirical counterparts computed on the CAMELYON16 dataset (i.e., $0.36$ and $0.06$, respectively). 
The class-specific means (i.e., $\mu_{1}$ and $\mu_{0}$) are set to be their sample counterparts computed on the instance-level features generated from CAMELYON16. The covariance matrix $\Sigma$ is treated similarly. The detailed parameter specification is included in Appendix B.3.
For computational simplicity, only the first $50$ dimensions of the feature vector are used. The detailed simulation results are presented in subsequent figures and their supporting raw data are given in Appendix B.4. Additional simulation experiments about various positive bag proportions and prediction results are presented in Appendix B.2. 

\begin{figure*}[!ht]
\centering
\subfigure{\includegraphics[scale=0.35]{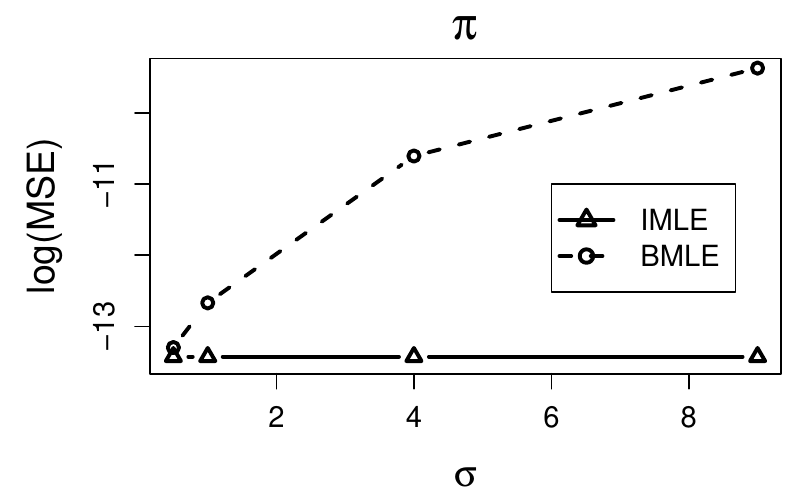}}
\hfil
\subfigure{\includegraphics[scale=0.35]{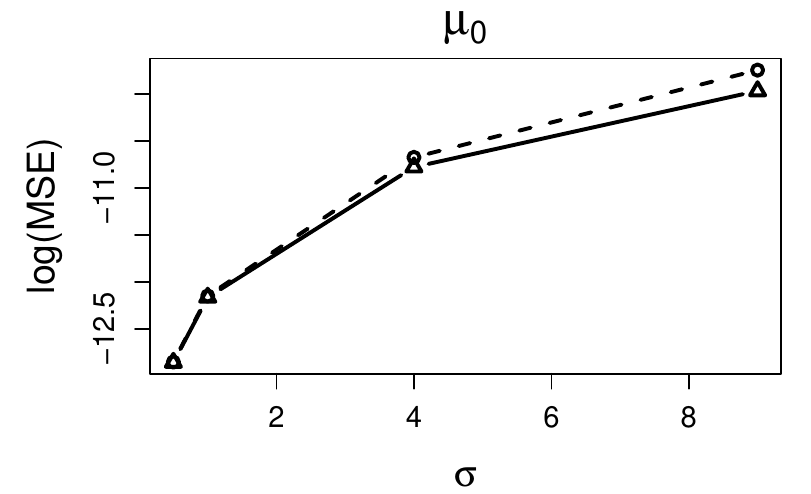}}
\hfil
\subfigure{\includegraphics[scale=0.35]{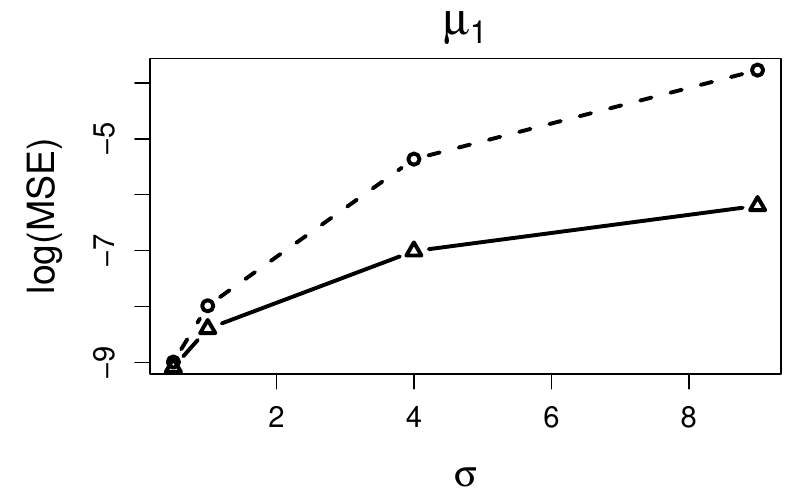}}
\hfil
\subfigure{\includegraphics[scale=0.35]{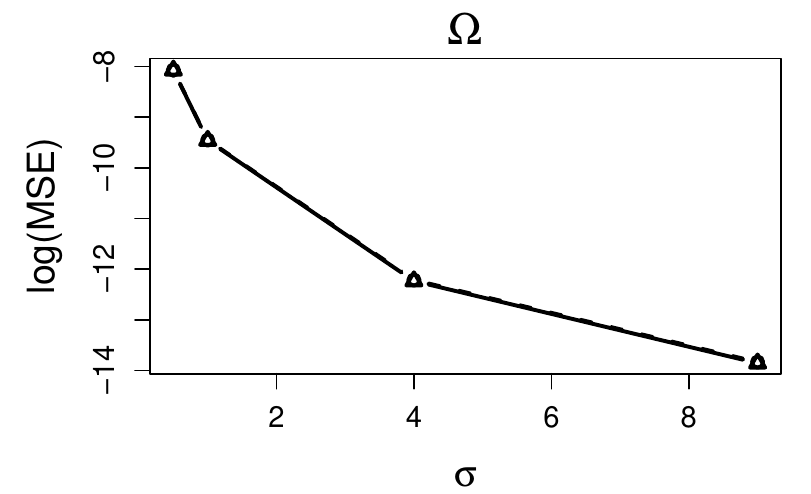}}
\caption{Log-transformed mean squared errors for the BMLE and the IMLE with different $\sigma$ values. The top left panel presents the estimation results for $\pi$. The bottom left panel is for $\mu_{1}$. The top right panel is for $\mu_{0}$. The bottom right panel is for $\Omega$. The solid line with triangles represents the IMLE. The dashed line with circles represents the BMLE.}
\label{Fig:Simu-1}
\end{figure*}

{\sc Study 1.} The objective of this study is to compare the statistical efficiency of the bag-based estimator $\widehat{\Theta}^{\bag}$ and the instance-based estimator $\widehat{\Theta}^{\ins}$. 
To this end, we fix the number of bags as $N=100$, the number of instances in each bag as $M=1{,}000$. We define $\Sigma_{\sigma} = \sigma \Sigma$ to be the covariance matrix for various $\sigma$ values in the data generating process.
For a given $\sigma$, we randomly replicate the experiment for a total of $R=500$ times. Let $\widehat{\Theta}^{(r)}$ be one particular type of estimator (e.g., $\widehat{\Theta}^{\bag}$) obtained in the $r$-th random replication. Write $\widehat{\Theta}^{(r)} = \big(\widehat{\pi}^{(r)}, \widehat{\mu}_{1}^{(r)\top}, \widehat{\mu}_{0}^{(r)\top}, \Vh(\widehat{\Omega}^{(r)})\big)\in\mathbb{R}^{q}$. 
For instance, we compute the mean squared error (MSE) for $\widehat{\mu}_{1}^{(r)}$ with $r=1,\cdots, R$ as $\MSE(\widehat{\mu}_{1}) = p^{-1}R^{-1}\sum_{r=1}^{R} \big\|\widehat{\mu}_{1}^{(r)} -\mu_{1}\big\|^{2}$.
It is then log-transformed and plotted in the top left panel of Figure \ref{Fig:Simu-1}. Similarly, the MSE values are also computed for $\widehat{\pi}^{(r)}$, $\widehat{\mu}_{0}^{(r)}$, and $\Vh(\widehat{\Omega}^{(r)})$. They are also log-transformed and plotted in different panels in Figure \ref{Fig:Simu-1}.

By the top left panel of Figure \ref{Fig:Simu-1}, we find that the instance-based estimator $\widehat{\pi}^{\ins}$ is statistically more efficient than the bag-based estimator $\widehat{\pi}^{\bag}$ in terms of $\log(\MSE)$. 
Moreover, the relative difference in terms of $\log(\MSE)$ increases as the level of noise $\sigma^2$ increases. 
Qualitatively similar pattern is also observed for $\mu_{1}$ from the bottom left panel. 
It is also interesting to note that the efficiency advantage of $\widehat{\mu}_{0}^{\ins}$ over $\widehat{\mu}_{0}^{\bag}$ is minimal as demonstrated in the top right panel of Figure \ref{Fig:Simu-1}. 
This finding is also expected since $\mu_{0}$ can be easily estimated by the instances from the negative bags only. Therefore, the efficiency improvement offered by the negative instances hidden in the positive bags is very limited. 
A qualitatively similar pattern is also observed for $\Omega$ in the bottom right panel of Figure \ref{Fig:Simu-1}.

\begin{figure*}[!ht]
\centering
\subfigure{\includegraphics[scale=0.35]{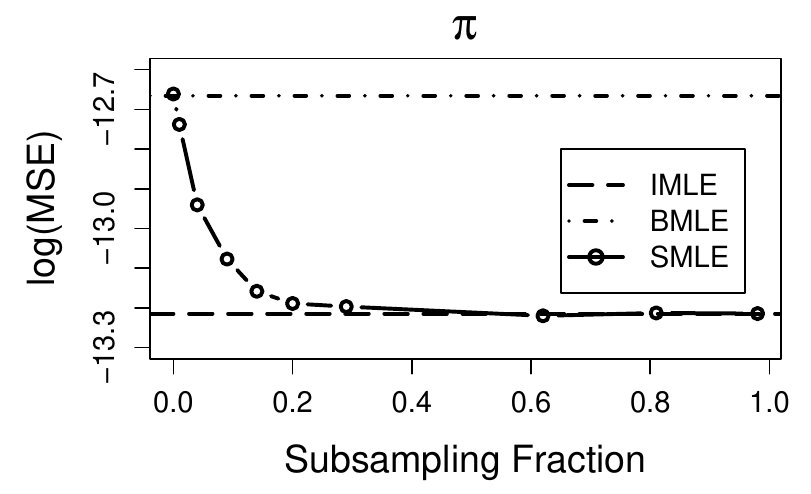}}
\hfil
\subfigure{\includegraphics[scale=0.35]{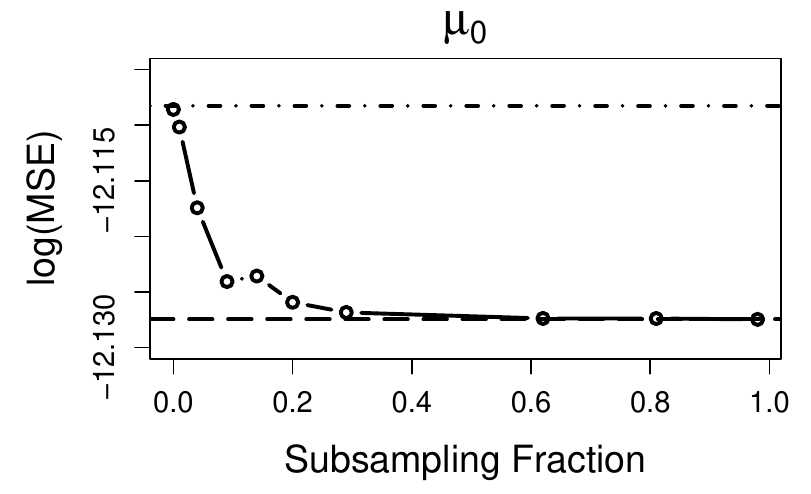}}
\hfil
\subfigure{\includegraphics[scale=0.35]{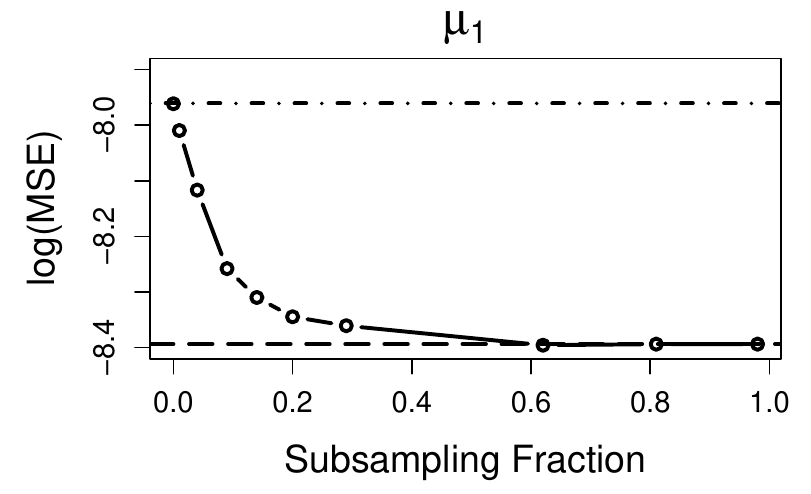}}
\hfil
\subfigure{\includegraphics[scale=0.35]{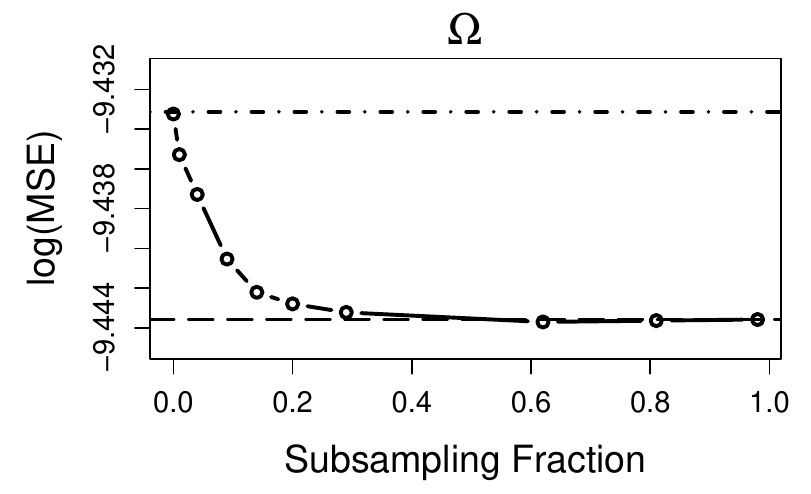}}
\caption{Log-transformed mean squared errors of the IMLE, the BMLE, and the SMLE with different subsampling fractions. The top left panel presents the estimation results for $\pi$. The bottom left panel is for $\mu_{1}$. The top right is for $\mu_{0}$. The bottom right is for $\Omega$. The dashed line represents the IMLE. The dotted line represents the BMLE. The solid line with circles represents the SMLE.}
\label{Fig:Simu-2}
\end{figure*}

{\sc Study 2.} The objective of this study is to illustrate the efficiency improvement of the subsampling-based estimator $\widehat{\Theta}^{\sub}$ with various subsampling fractions. More specifically, we adopt a very similar parameter setting as the previous study. The only difference is that the covariance matrix is fixed to be $\Sigma$. We next vary the $\alpha_{n}$ value in \eqref{sampling-prob}. This leads to different subsampling fractions from $0\%$ to $100\%$. 
We then replicate the experiment in a similar way as in the {\sc Study 1}. The detailed results are summarized in Figure \ref{Fig:Simu-2}. The key difference is that the horizontal axis is changed from $\sigma^2$ in Figure \ref{Fig:Simu-1} to the subsampling fraction in Figure \ref{Fig:Simu-2}. For comparison purpose, the results of both IMLE and BMLE are also presented. 
Both these estimators are not affected by the subsampling fraction. 
Therefore, the corresponding $\log(\MSE)$ curves in Figure \ref{Fig:Simu-1} become two horizontal reference lines in Figure \ref{Fig:Simu-2}. 
They are obviously separated for $\pi$ (the top left panel) and $\mu_{1}$ (the bottom left panel). 
For $\mu_{0}$ (the top right panel) and $\Omega$ (the bottom right panel), these two horizontal reference lines are close to each other. 
These findings are in accordance with those of the {\sc Study 1}.

By Figure \ref{Fig:Simu-2}, we find that the estimating accuracy of the SMLE for $\Theta$ is quite similar to that of the BMLE, if the subsampling fraction is low (e.g., $<1\%$). 
This is an expected result. 
With such a low subsampling fraction, the total number of instances utilized by the SMLE is also small. 
Therefore, the only information used in the estimation is the bag-level information. 
However, as the subsampling fraction increases, we find that the $\log(\MSE)$ value of the SMLE for $\pi$ becomes smaller than that of the BMLE. The advantage steadily improves as the subsampling fraction increases. This suggests that, with a reasonably large subsampling fraction, the SMLE can estimate every component of $\Theta$ much better than the BMLE. If the subsampling fraction is large enough, the SMLE can perform as well as the IMLE. In this case, the SMLE becomes the IMLE.

\begin{figure*}[!ht]
\centering
\subfigure{\includegraphics[scale=0.45]{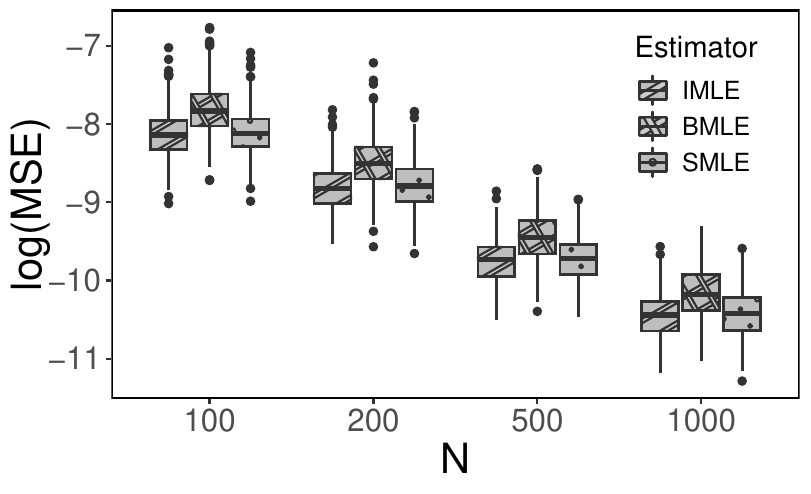}}
\hfil
\subfigure{\includegraphics[scale=0.45]{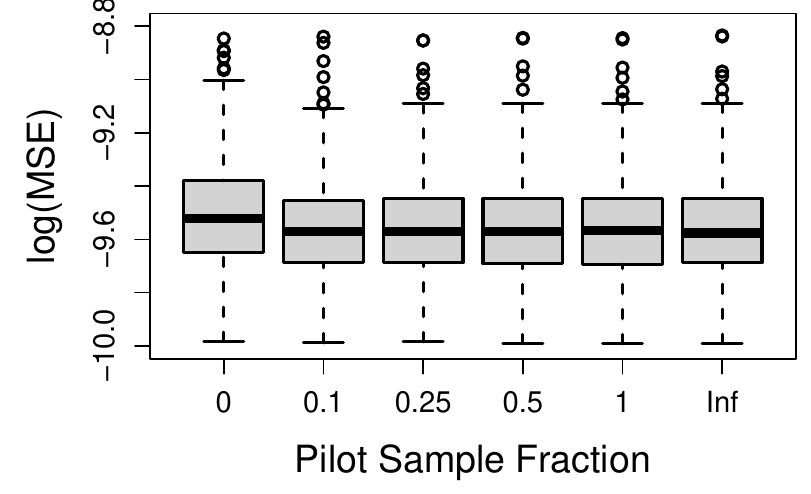}}
\caption{The boxplot of $\log(\MSE)$ values for different estimators. The left panel illustrates the estimation results with different sample sizes. The right panel illustrates the estimation results for the SMLE with different pilot sample sizes.}
\label{Fig:Simu-4}
\end{figure*}

{\sc Study 3.} Next, we want to focus on the statistical consistency of the proposed estimators. 
The experiment is conducted in a similar way as before but with a different parameter variation strategy. The left panel of Figure \ref{Fig:Simu-4}  provides the $\log(\MSE)$ results with different sample sizes for different estimators, while all the other parameters are fixed as in the {\sc Study 2}. 
By the left panel of Figure \ref{Fig:Simu-4}, we find that the $\log(\MSE)$ values for all the three estimators steadily decrease as the sample size $N$ increases. This confirms the fact that all the three estimators, i.e., IMLE, BMLE, and SMLE, are all statistically consistent with the same convergence rate. 

{\sc Study 4.} 
The objective of the last study is to investigate the uncertainty effect of the pilot estimator $\widehat{\beta}$ in determining the asymptotic efficiency of $\widehat{\Theta}^{\sub}$. To this end, we need $\widehat{\beta}$s to be of different uncertainty levels. 
This can be done by varying the pilot sample sizes used for $\widehat{\beta}$ estimation. The fraction of this pilot sample varies from $0\%$ to $\infty$, where $0\%$ here refers to the uniform subsampling without estimating $\beta$, and ``Inf" ($\infty$) refers to using the true parameter $\beta = \Omega(\mu_{1}-\mu_{0})$ for subsampling probability specification. We then replicate the experiment in a similar way as in the {\sc Study 3}. The MSEs of $\widehat{\Theta}^{\sub}$ are log-transformed and box-plotted in the right panel of Figure \ref{Fig:Simu-4}.
By the right panel of Figure \ref{Fig:Simu-4}, we find that the $\log(\MSE)$s with different pilot sample sizes are similar, except for the case with $0\%$ pilot sample fraction. 
This confirms the fact that the uncertainty in $\widehat{\beta}$ plays an asymptotically ignorable role in determining the asymptotic efficiency of $\widehat{\Theta}^{\sub}$, as long as $\widehat{\beta}$ is statistically consistent. This corroborates our asymptotic results in Section \ref{sec:method-smle} very well. Therefore, practitioners can use essentially any consistent pilot estimator, as long as their estimated standard errors are reasonably small.

\section{Robustness Analysis}\label{sec:robust}

Despite the interesting findings obtained in the previous sections, it is of great interest to extend our method to adapt to more general settings. We systematically evaluate the robustness of our method under potential model mis-specifications and demonstrate its broad applicability across three different scenarios. 
To save space, we focus on the robustness properties of the BMLE. Those of the IMLE and SMLE can be treated similarly and are validated numerically.

\subsection{Heterogeneous Mixing Probability $\pi_{i}$}\label{sec:robust-heteromixing}

We start with the homogeneous assumption about the instance-level mixing probability $\pi$. Specifically, our model \eqref{log-like} assumes that different bags share a common mixing probability $\pi$. 
As we mentioned before in Section \ref{sec:datades}, the percentage of sub-images related to tumor on each positive WSI varies greatly with a mean of 6.61\% and a standard deviation of 24.85\%.
Instead, it is theoretically more appealing to assume for different bag $i$ a different mixing probability $\pi_{i}$, which behaves like a bag-specific mixed effect and should be independently generated from some unknown distribution. It is then of great interest to evaluate the robustness of the BMLE against this model mis-specification.

More specifically, we assume that $\pi_{i}$ with different $i$s are independently generated from some unknown distribution $g(\cdot)$ with $E(\pi_{i})=\pi$. Then, we are interested in studying the statistical consistency of $\widehat{\Theta}^{\bag}$, which is computed by maximizing the incorrectly specified log-likelihood function $\mathcal{L}^{\bag}\big(\Theta\big)$.
Note that this is a log-likelihood function with bag-specific mixing probabilities $\pi_{i}$s incorrectly specified as a common value $\pi$. It is remarkable that if different bags indeed share the same mixing probability $\pi$, we indeed should have $(NM)^{-1}E\big\{\dot{\mathcal{L}}^{\bag}(\Theta)\big\} = 0$ exactly. 
In fact, this is the key condition responsible for the statistical consistency of $\widehat{\Theta}^{\bag}$ under the homogeneous mixing probability assumption \citep{van2000asymptotic}. 
However, with the bag-specific mixing probability $\pi_{i}$, whether we remain to have $(NM)^{-1}E\big\{\dot{\mathcal{L}}^{\bag}(\Theta)\big\} = 0$ is not immediately clear. 
To this end, we need to re-study this condition of the mis-specified log-likelihood function \eqref{log-like} but under the assumption that the true bag-level mixing probabilities are heterogeneous. 
Interestingly, it can be verified that
\begin{gather}
E\big\{\dot{\mathcal{L}}^{\bag}(\Theta)\big\} = 0 , 
\label{eq: sec4.1 E} \\
(NM)^{-1}\cov\big\{\dot{\mathcal{L}}^{\bag}(\Theta)\big\} = \Omega^{\bag}. 
\label{eq: sec4.1 cov}
\end{gather}
The verification details of \eqref{eq: sec4.1 E} and \eqref{eq: sec4.1 cov} are given in Appendix D.1.
Then by similar and additional arguments as in \cite{fan2001variable}, we can show that $\widehat{\Theta}^{\bag}-\Theta = -\big\{\ddot{\mathcal{L}}^{\bag}(\Theta)\big\}^{-1} \dot{\mathcal{L}}^{\bag}(\Theta) \big\{1+o_{p}(1)\big\} = O_{p}\big(1/\sqrt{NM}\big)$, even if the true mixing probability $\pi_{i}$ is heterogeneous. 
This suggests that the BMLE remains to estimate $\mu_k$ and $\Omega$ consistently.

\begin{figure*}[!ht]
\centering
\subfigure{\includegraphics[scale=0.24]{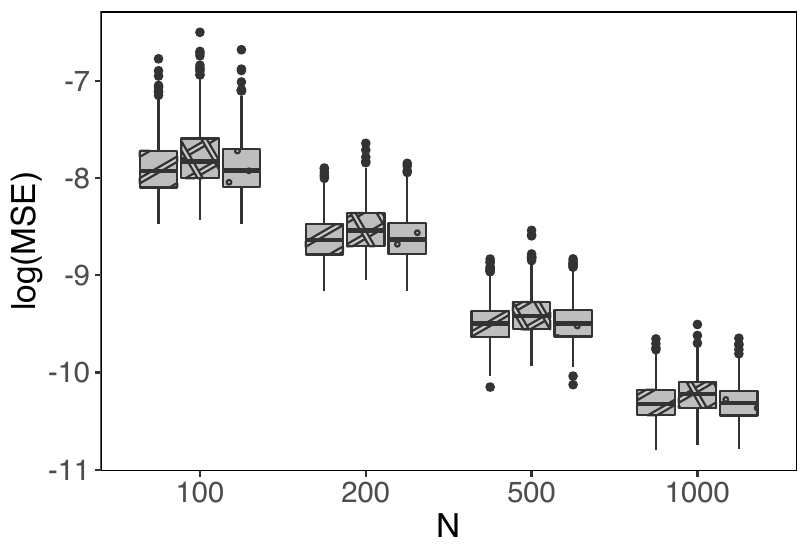}}
\hfil
\subfigure{\includegraphics[scale=0.24]{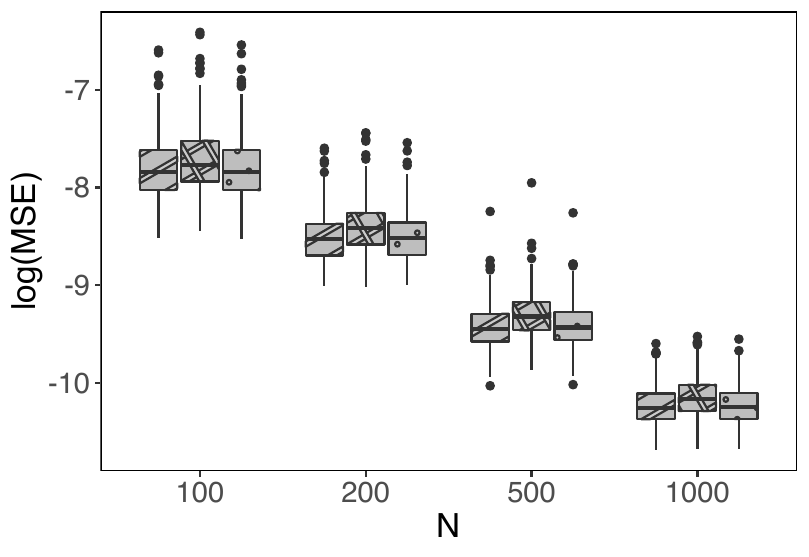}}
\hfil
\subfigure{\includegraphics[scale=0.225]{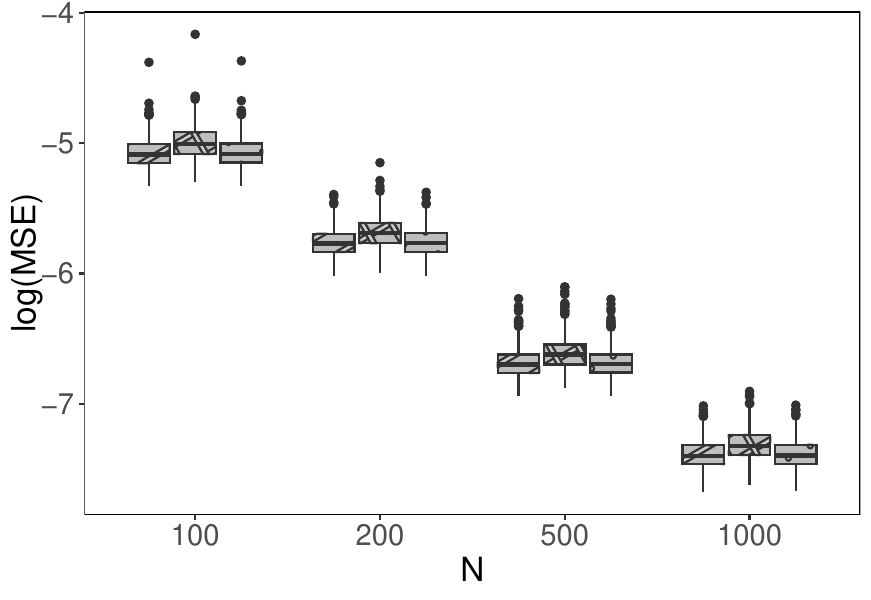}}
\hfil
\subfigure{\includegraphics[scale=0.225]{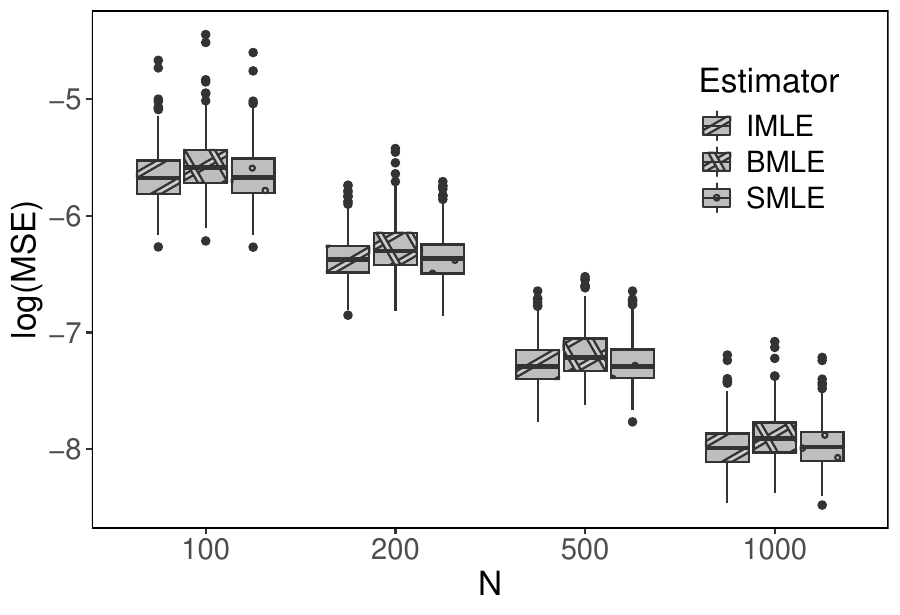}}
\caption{Log-transformed mean squared errors of key parameters for the proposed estimators with different $N$ values under various model mis-specification. The left panel presents the estimation results under the heterogeneous mixing probability $\pi_{i}$. The middle left panel presents the estimation results under the spatial correlation of $A_{im}$. The middle right panel presents the estimation results under the spatial correlation of $X_{im}$. The right panel presents the estimation results under the conditional dependence of $A_{im}$.}
\label{Fig:RealSimu}
\end{figure*}

To numerically verify the above theoretical claim, we slightly modify the setting of simulation {\sc Study 3} so that bag-specific mixing probability is allowed. Specifically, different $\pi_{i}$s are independently generated. For a given $i$, we randomly select a WSI sample for the CAMELYON16 dataset. We next compute the empirical proportion of the positive instances for this WSI sample. It is then set to be the empirical distribution of $\pi_{i}$. Other than $\pi_{i}$s, all the simulation parameters (e.g., $N$, $M$, $\mu_{k}$s) are set to be the same as those in {\sc Study 3}. Moreover, for comparison purposes, all the three estimators (i.e., the IMLE, BMLE, and SMLE) are computed to validate the robustness. 
The experiment results are then summarized in the same way as before in the left panel of Figure \ref{Fig:RealSimu}. 
We find that the IMLE, BMLE, and SMLE are all statistically consistent for estimating the key parameters (i.e., $\mu_{k}$s and $\Omega$). 
This implies that our methods are fairly robust, in the sense that even if heterogeneous mixing probability $\pi_{i}$ is assumed, the resulting estimators remain statistically consistent. 

\subsection{The Spatial Independence Assumption for $A_{im}$}\label{sec:robust-spatial}

Another limitation of our model is that it implicitly assumes that different positive instances in the same bag are randomly distributed across different locations. 
Therefore, we should not expect positive instances to be clustered spatially. 
Unfortunately, this is not the case in real practice. 
In real practice, the positive instances (i.e., the metastasis) are often clustered spatially \citep{yu2025semiparametric}; see Figure \ref{Fig:TissueMask} (c) for some understanding. 
It is then of great interest to evaluate the robustness of our method against this plausible assumption violation. 

To this end, we follow \cite{diggle2013statistical} and assume for each instance $(i,m)$ a spatial location $R_{im}\in\mathbb{R}^{2}$. 
We next assume that $P(A_{im}=1|R_{im}=r,Y_{i}=1) = \pi_{i}(r)$, where $\pi_{i}(r)$ is the bag-specific, spatially varying, and smooth local mixing probability with $r\in\mathbb{R}^{2}$. 
With a slight abuse of notation, we redefine $\pi = E\big\{\pi_i(R_{im})\big\}$, which is used in the mis-specified log-likelihood function \eqref{log-like}. 
Under this model setup, those locations with large $\pi_{i}(r)$ values should be locally clustered \citep{diggle2013statistical}. 
This naturally leads to locally clustered positive instances. 
However, with the spatial dependent $A_{im}$s, whether we remain to have $(NM)^{-1}E\big\{\dot{\mathcal{L}}^{\bag}(\Theta)\big\} = 0$ is not immediately clear. 
To this end, we need to re-study this condition of the mis-specified log-likelihood function \eqref{log-like} but under the spatial dependence assumption. 
By similar arguments as in the previous Section \ref{sec:robust-heteromixing}, we can rigorously verify that both \eqref{eq: sec4.1 E} and \eqref{eq: sec4.1 cov} remain valid.
The rigorous theoretical verification details of those two conclusions are given in Appendix D.2.
This implies that $\widehat{\Theta}^{\bag}-\Theta = - \big\{\ddot{\mathcal{L}}^{\bag}(\Theta)\big\}^{-1} \dot{\mathcal{L}}^{\bag}(\Theta)\big\{1+o_{p}(1)\big\} = O_{p}\big(1/\sqrt{NM}\big)$ with the bag-specific and spatially varying mixing probability $\pi_{i}(R_{im})$. 
Therefore, we are able to show that the BMLE remains to estimate $\mu_{k}$s and $\Sigma$ consistently, even if it is developed with a mis-specified spatially non-varying mixing probability $\pi$.

To numerically verify this theoretical claim, we slightly modify the setting of simulation {\sc Study 3} again so that the spatial correlation is allowed. Specifically, the spatial location $R_{im}$ is randomly generated in a unit circle. Then, we set $\pi_{i}(r) = \pi_{i}\times p(r)$, where $p(r)=\sin(r_{1})\times \sin(r_{2})$ for any $r\in \mathbb{R}^{2}$ and is normalized to be $p(r)\in[0.8,1]$. All the other simulation parameters (e.g., $N$, $M$, $\mu_{k}$s) are set to be the same as those in {\sc Study 3}. The experiment is then replicated and summarized as before in the middle left panel of Figure \ref{Fig:RealSimu}. 
We find that the IMLE, BMLE, and SMLE are all statistically consistent for estimating the key parameters (i.e., $\mu_{k}$s and $\Omega$). This implies that our method is fairly robust against the plausible violation of the spatial independence assumption for $A_{im}$.

\subsection{The Spatial Independence Assumption for $X_{im}$}\label{sec:robust-X-spatial}

In real practice, we often have the feature vector $X_{im}$s spatially dependent, since they are collected from different locations on the WSI. With spatially dependent $X_{im}$s, our model defined in Section \ref{sec:method-imle} becomes mis-specified. It is then of interest to study the robustness of our method against this potential model mis-specification. To this end, we rewrite $X_{im}$ as $X_{im} = \mu_{A_{im}} + \epsilon_{i}(R_{im})$, where $R_{im}\in\mathbb{R}^{2}$ is the spatial location at which $X_{im}$ is sampled. Next, we assume for the random noise $\epsilon_{i}(s)\in\mathbb{R}^{p}$ with $s\in\mathbb{R}^{2}$ a bag-specific Gaussian random field with $E\big\{\epsilon_{i}(s)\big\} = 0$ and $\cov\big\{\epsilon_{i}(s_{1}), \epsilon_{i}(s_{2})\big\} = \exp\big(-\|s_{1} - s_{2}\|/\phi\big)\Sigma$, where $\phi>0$ controls the spatial correlation strength. It follows then $\cov\big\{\epsilon_{i}(s)\big\}=\Sigma$. Accordingly, the spatial dependence among different $X_{im}$s on the same slide $i$ can be accommodated. With $\phi \rightarrow 0$, this model misspecification reduces to our Gaussian mixture model defined in Section \ref{sec:method-imle}. 
Next, let $\widehat{\Theta}^{\bag}$ be the maximizer of \eqref{log-like} computed on those spatially dependent $X_{im}$s but without taking their spatial dependence structure into consideration.
In this case, $\widehat{\Theta}^{\bag}$ is no longer the genuine MLE. Instead, it is a pseudo maximum likelihood estimator \citep[PMLE,][]{besag1986statistical} based on a pseudo log-likelihood function. Then, whether this PMLE $\widehat{\Theta}^{\bag}$ remains statistically consistent becomes an interesting problem. 
We then have the following Theorem \ref{thm-robust-X}.
\begin{theorem}\label{thm-robust-X}
We have (1) $\|\widehat{\Theta}^{\bag} - \Theta\| = O_{p}(1/\sqrt{NM})$ and (2) $\sqrt{NM}\big(\widehat{\Theta}^{\bag} - \Theta\big)\rightarrow_{d}N\big\{0,\big(\Omega^{\bag}\big)^{-1}\widetilde{\Omega}^{\bag}\big(\Omega^{\bag}\big)^{-1}\big\}$ as $\min(N, M)\rightarrow\infty$.
\end{theorem}
The detailed proof of Theorem \ref{thm-robust-X} and the definition of $\widetilde{\Omega}^{\bag}$ is provided in
Appendix D.4. Comparing Theorem \ref{thm-robust-X} and Theorem \ref{thm-2}, we find that the PMLE $\widehat{\Theta}^{\bag}$ remains $\sqrt{NM}$-consistent and asymptotically normal. The only difference is the asymptotic covariance. 
To numerically verify this theoretical claim, we modify the setting of simulation {\sc Study 3} slightly so that spatial dependence can be allowed in the instance-level features. Specifically, the spatial location $R_{im}$ is randomly generated over a disk with radius $r = c\sqrt{M}$ and $c=0.03$. Next, for each bag $i$, we then generate $\{X_{im}\}_{m=1}^M$ from the Gaussian random field with correlation range $\phi=0.3$.
The experiment is then replicated and summarized as before in the middle right panel of Figure \ref{Fig:RealSimu}. 
We find that the IMLE, BMLE, and SMLE all remain statistically consistent, which corroborates our theoretical claims of Theorem \ref{thm-robust-X} very much. Therefore, we are able to show that the proposed estimators remain statistically valid, even though the generating feature model is spatially dependent.

\subsection{Conditional Independence Assumption about $A_{im}$}\label{sec:robust-conditional}

Another assumption used by our method is the conditional independence assumption, which assumes that different $A_{im}$s with the same $i$ but different $m$ are mutually independent conditional on $Y_{i}=1$. 
This leads to
\begin{gather}
P\Big(A_{i}=v|Y_{i}=1\Big) = \prod\limits_{m=1}^{M} \pi^{v_{m}}\Big(1-\pi\Big)^{1-v_{m}},
\label{Aim-dep-previous}
\end{gather}
where $A_{i}=(A_{i1},\cdots,A_{iM})^{\top}\in\{0,1\}^{M}$ and $v_{m}\in\{0,1\}$ for every $1\leq m\leq M$. Practically, this assumption can be easily violated as follows. Note that for a positive WSI sample with $Y_{i}=1$, we must have $\sum_{m=1}^{M} A_{im}>0$. 
However, our conditional independence assumption about $A_{im}$ allows $\sum_{m=1}^{M} A_{im} =0$ with a small probability as $(1-\pi)^M$ even if $Y_{i}=1$. 
To fix this problem, we can slightly modify \eqref{Aim-dep-previous} to be $P\big(A_{i}=v|Y_{i}=1\big)= C_{A}^{-1}\prod_{m=1}^{M} \pi^{v_{m}}\big(1-\pi\big)^{1-v_{m}}$ but under the constraint that $\sum_{m=1}^{M}v_{m}>0$. Here $C_{A} = 1-(1-\pi)^{M}$ is the normalizing constant. We then have
\begin{gather}
P\Big(A_{i}=v|Y_{i}=1\Big) = \Big\{1-(1-\pi)^{M}\Big\}^{-1}\prod\limits_{m=1}^{M} \pi^{v_{m}}\Big(1-\pi\Big)^{1-v_{m}},
\label{Aim-dep}
\end{gather}
where $v\in\mathcal{V} = \big\{v = (v_{m}): \sum_{m=1}^{M} v_{m}>0 \text{ with } v_{m}\in\{0,1\}\big\}$. 
As one can see, the probability distribution \eqref{Aim-dep} is similar to the probability distribution \eqref{Aim-dep-previous}. 
The only difference is that $\sum_{m=1}^{M} A_{im}$ is constrained to be positive for any $A_{i}$ generated according to \eqref{Aim-dep}. 
The price paid for this theoretical rigidity is that $A_{im}$s with the same $i$ but different $m$s are no longer conditionally independent. 

More specifically, conditional on $A_{i}=v$ for some $v\in\mathcal{V}$, we have $\V(\mathbb{X}_{i})\in\mathbb{R}^{Mp}$ follows a multivariate Gaussian distribution with mean $v\otimes\mu_{1} + (1_{M}-v)\otimes\mu_{0}$ and covariance 
$\diag(vv^\top)\otimes \Sigma + \{I_{M}-\diag(vv^\top)\} \otimes \Sigma$, where $1_M = (1,\dots,1)^\top \in \mathbb{R}^{M}$, and $I_M \in \mathbb{R}^{M\times M}$ is an identity matrix. 
Then, the bag-based log-likelihood in \eqref{log-like} in this case becomes
\begin{gather*}
\mathcal{L}_{*}^{\bag}(\Theta) = \sum\limits_{i=1}^{N}\Big(1-Y_{i}\Big) \sum\limits_{m=1}^{M} \log \phi_{\mu_{0},\Sigma} \Big(X_{im}\Big) \\
+ \sum\limits_{i=1}^{N}Y_{i} \log\bigg[\Big\{1-(1-\pi)^{M}\Big\}^{-1} \sum\limits_{v_{i}\in \mathcal{V}} \Big\{\prod\limits_{m=1}^{M} \pi^{v_{im}}\big(1-\pi\big)^{1-v_{im}} \phi_{\mu_{v_{im}},\Sigma_{v_{im}}}\big(X_{im}\big)\Big\} \bigg],
\end{gather*}
which seems to be very different from that in \eqref{log-like}. 
However, since $\prod_{m=1}^{M}\big\{\pi\phi_{\mu_{1},\Sigma}(X_{im})$ 
$+ (1-\pi) \phi_{\mu_{0},\Sigma}(X_{im})\big\} 
= \sum_{v_{i}\in \mathcal{V}} \big\{\prod_{m=1}^{M} \pi^{v_{im}}(1-\pi)^{1-v_{im}} \phi_{\mu_{v_{im}},\Sigma_{v_{im}}}(X_{im})\big\} + \big(1-\pi\big)^{M}$ 
$\prod_{m=1}^{M} \phi_{\mu_{0},\Sigma}\big(X_{im}\big)$, 
it follows then 
\begin{gather}
\mathcal{L}_{*}^{\bag}(\Theta) = \sum\limits_{i=1}^{N}\Big(1-Y_{i}\Big)\sum\limits_{m=1}^{M}\log \phi_{\mu_{0},\Sigma}\Big(X_{im}\Big)
+ \sum\limits_{i=1}^{N}Y_{i}\Bigg[-\log\bigg\{1-\Big(1-\pi\Big)^{M}\bigg\} + \notag
\\ \log\bigg\{\prod\limits_{m=1}^{M}\Big[\pi\phi_{\mu_{1},\Sigma}\big(X_{im}\big) + \big(1-\pi\big) \phi_{\mu_{0},\Sigma}\big(X_{im}\big)\Big] -\Big(1-\pi\Big)^{M}\prod\limits_{m=1}^{M}\phi_{\mu_{0},\Sigma}\Big(X_{im}\Big)\bigg\}\Bigg].
\label{dep-bag-like}
\end{gather}
Comparing \eqref{dep-bag-like} with \eqref{log-like}, we find that both the log-likelihood functions are similar since $(1-\pi)^{M}\rightarrow 0$ as $M\rightarrow\infty$.
Therefore, one can reasonably expect that the BMLE remains consistent under probability distribution \eqref{Aim-dep}. This can be rigorously explained as follows.

Recall under the assumption that different $A_{im}$s conditional on $Y_{i}=1$ are indeed independently generated, we should have $(NM)^{-1}E\big\{\dot{\mathcal{L}}^{\bag}(\Theta)\big\} = 0$ exactly. 
Once again, we need to study this condition of the mis-specified log-likelihood function \eqref{log-like} but with $A_{im}$s generated according to the probability distribution as \eqref{Aim-dep}. 
Interestingly, it can be verified that
\begin{gather}
(NM)^{-1}E\big\{\dot{\mathcal{L}}^{\bag}(\Theta)\big\} = O\big\{(1-\pi)^{M}\big\}, \label{eq: sec4.3 E} \\
(NM)^{-1}\cov\big\{\dot{\mathcal{L}}^{\bag}(\Theta)\big\} = \Omega^{\bag}\big\{1+o(1)\big\}. 
\label{eq: sec4.3 cov}
\end{gather}
The verification details of \eqref{eq: sec4.3 E} and \eqref{eq: sec4.3 cov} are given in Appendix D.3.
Then we can show that $\widehat{\Theta}^{\bag}-\Theta = - \big\{\ddot{\mathcal{L}}^{\bag}(\Theta)\big\}^{-1}\dot{\mathcal{L}}^{\bag}(\Theta)$
$\big\{1+o_{p}(1)\big\} = O_{p}\big\{1/\sqrt{NM} + (1-\pi)^{M}\big\}$ under the true data generating distribution \eqref{Aim-dep} when $\min(N,M)\rightarrow\infty$. Therefore, we have that as long as $\log N/M\rightarrow 0$, we should have $\sqrt{NM}(1-\pi)^{M}\rightarrow 0$ as $M\rightarrow\infty$. To gain some practical understanding about these conditions, we take the CAMELYON16 dataset as an example, where we have $\pi = 6.6\%$, $N=247$ and $M=10{,}000$. We then have $(1-\pi)^{M} = 2.9\times 10^{-297}$ and $\log N/M = 5.5\times 10^{-4}$. Thus, both the conditions above seem to be well satisfied. In this case, we remain to have $\widehat{\Theta}^{\bag}-\Theta = O_{p}(1/\sqrt{NM})$. 
This suggests that the BMLE is fairly robust against the plausible violation of the conditional independence assumption, in the sense that the resulting estimator remains statistically consistent.

To numerically verify this theoretical claim, we slightly modify the setting of simulation {\sc Study 3} again so that the conditional independence assumption about $A_{im}$ is violated. Specifically, we generate $A_{im}$s according to \eqref{Aim-dep} so that $\sum_{m=1}^{M} A_{im}>0$ is always satisfied when $Y_{i}=1$. All the other simulation parameters (e.g., $N$, $M$, $\mu_{k}$s) are set to be the same as those in {\sc Study 3}. 
The experiment is then replicated and summarized as before in the right panel of Figure \ref{Fig:RealSimu}. 
We find that the IMLE, BMLE, and SMLE all remain statistically consistent for the key parameters (i.e., $\mu_{k}$s and $\Omega$). 
This implies that our method is fairly robust against the plausible violation of the conditional independence assumption about $A_{im}$.

\section{Application to the CAMELYON16 Data}\label{sec:realdata}

\subsection{Data Preprocessing}\label{sec:datapre}

\begin{figure}[!htbp]
\centering
\includegraphics[scale=0.55]{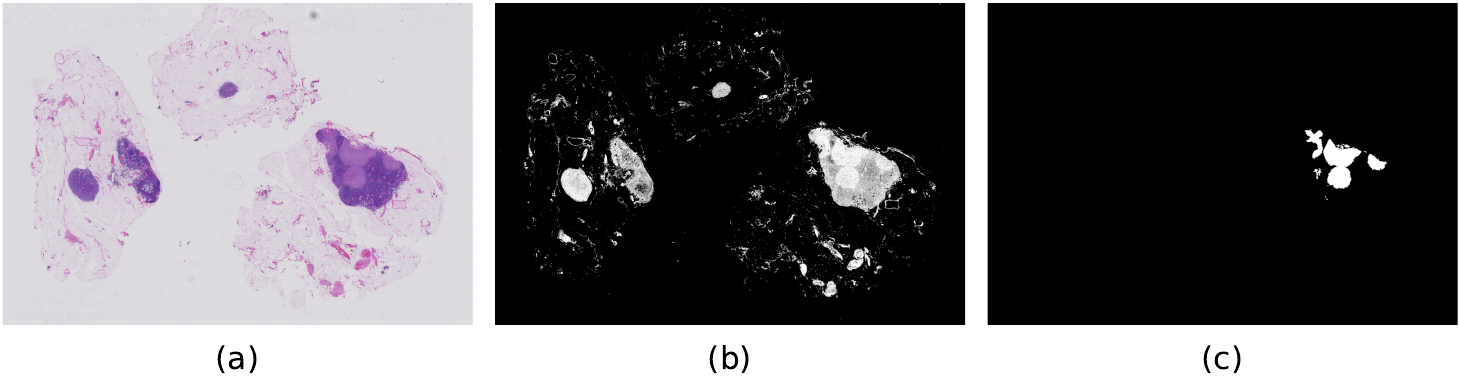}
\caption{The illustration of a WSI sample used in the training data. The left panel (a) is the original image. The bright area in the middle panel (b) corresponds to the pixels with tissue information. The bright area in the right panel (c) corresponds to the pixels with metastatic carcinoma.}
\label{Fig:TissueMask}
\end{figure}

We next apply our method to the CAMELYON16 dataset of \cite{bejnordi2017diagnostic}. We first consider how to generate instances from the data. 
Note that each WSI varies in size, and the sizes are generally very large across different samples. 
Therefore, it is very difficult to directly treat the original WSI as a whole. One practical solution as advocated by \cite{wang2016deep} is to generate a large number of randomly sampled sub-images with smaller but identical sizes. These sub-images are then treated as the instances contained in the bag. To this end, we first apply the algorithm of \cite{otsu1979threshold} to exclude those pixels related with the background region for each WSI sample. 
These pixels are often associated with the paraffin wax used in pathology slides. 
Therefore, they are totally useless for diagnosis; see the gray background areas in the left panel of Figure \ref{Fig:TissueMask}. 
Thereafter, we should focus on those remaining and more informative pixels; see the bright area in the right panel of Figure \ref{Fig:TissueMask}. 
Next, we randomly select a pixel from the informative region, i.e., the bright area in Figure \ref{Fig:TissueMask}(b). 
Take this sampled pixel as the center position and crop sub-images of various sizes. 
Our experiment results in Appendix B.1 suggest that larger sub-image sizes lead to better predictive performances, since more information is contained in the sub-images with larger sizes.
However, given the fact that the graphics memory sizes of our GPUs (i.e., NVIDIA A30 GPUs with a total of $7\times24=168$ GB graphics memory) are limited, we cannot process sub-images with too large sizes either.
This leads to the final sub-image size of $512\times 512$ for not only its outstanding predictive performance on the CAMELYON16 dataset but also its practical feasibility.
Next, we define $A_{im} =1$ if the selected sub-image happens to be related to the metastatic region, and define $A_{im}=0$ otherwise.

Having generated the instances (i.e., sub-images) from the WSIs, the next task is to extract the features from these sub-images for metastasis detection. To convert each sub-image into a feature vector, various deep learning models have been considered. Those deep learning models include but are not limited to the VGG of \cite{simonyan2014very}, ResNet of \cite{he2016deep}, and ShuffleNet of \cite{zhang2018shufflenet}. 
In our case, we choose to apply the UNI model of \cite{chen2024towards} since it is one of the SOTA general-purpose foundation models for pathology. It is based on the Vision Transformer architecture of \cite{dosovitskiy2020image}, and has been pre-trained using the DINOv2 method \citep{oquab2023dinov2} on more than $10^8$ images from over $10^5$ diagnostic H\&E-stained WSIs with over $77$ TB data. Therefore, we are motivated to apply the pre-trained UNI model to the sub-images (i.e., instances). This leads to a feature vector $X_{im}$ of $p=512$ dimension for each sub-image instance. To gain a quick understanding of how $X_{im}$s differ across instances with and without metastases, see Appendix B.1 for a graphical illustration. {To partially deal with the high dimensionality issue, PCA-based analyses are experimented and illustrated in Appendix B.1.}

\subsection{Baseline Methods}\label{sec:baseline}

{In this subsection, we introduce a number of competing baseline methods that are then compared with our method. The first method is a UNI-based ABMIL method as reported by \cite{chen2024towards}, which applies the ABMIL model of \cite{ilse2018attention} for weakly supervised classification. The second method is the ItS2CLR (Iterative Self-paced Supervised Contrastive Learning for MIL Representations) model of \cite{liu2023multiple}, which updates the feature extracting network by contrastive learning. The third method is the TransMIL method of \cite{shao2021transmil}, which encodes the spatial information of WSIs in the attention mechanism. The fourth method is the IRDTV method of \cite{wu2024individualized}, which is a deep logistic model with total variation regularization. The fifth one is the PANTHER of \cite{song2024morphological}, which utilizes GMM to learn task-agnostic prototypes in a fully unsupervised way.} 

\subsection{Predicting Results}\label{sec:pred}

{With the help of the feature extracted by the UNI model, our method can be readily applied. Specifically, we compute on the training data a total of three estimators. They are, respectively, the IMLE, the BMLE, and the SMLE. For the SMLE method, the subsampling fraction is fixed to be $10\%$. Experiments with different subsampling fractions are to be illustrated in Appendix B.1.}

To gauge the predictive performance, two levels of measures are considered. They are bag-level measures and instance-level measures.
Consider an arbitrary testing bag $i$ with a bag-level label $Y_{i}^{\test}$. 
Next compute for it the estimated bag-level positive probability, which is given by $\widehat{Y}_{i}^{\test} = 1 - \prod_{m=1}^{M} \big(1-\widehat{\pi}_{im}^{\test}\big)$. 
Here $\widehat{\pi}_{im}^{\test} = \exp\big(\widehat{\alpha}_{0}^{*}+X_{im}^{\test\top}\widehat{\beta}\big)\big/\big\{1+\exp(\widehat{\alpha}_{0}^{*}+X_{im}^{\test\top}\widehat{\beta})\big\}$ is the predicted positive probability of the $m$-th instance for the $i$-th test bag sample, where $\widehat{\alpha}_{0}^{*}=\big(\widehat{\mu}_{0}^{*\top}\widehat{\Omega}^{*}\widehat{\mu}_{0}^{*} - \widehat{\mu}_{1}^{*\top}\widehat{\Omega}^{*}\widehat{\mu}_{1}^{*}\big)\big/2 + \log\big\{(1-\widehat{\pi}^{*})^{-1}\widehat{\pi}^{*}\big\}$, $\widehat{\beta} = \widehat{\Omega}^{*}(\widehat{\mu}_{1}^{*}-\widehat{\mu}_{0}^{*})$, and $\widehat{\Theta}^{*} = \big(\widehat{\pi}^{*}, \widehat{\mu}_{1}^{*\top}, \widehat{\mu}_{0}^{*\top}, \Vh(\widehat{\Omega}^{*})^{\top}\big)^{\top}$ represents one particular estimator (e.g., $\widehat{\Theta}^{\sub}$). 
This leads to a total of $128$ prediction results. We can then compute bag-level AUC, AUPRC, F1 score, Recall, and Precision values by using $(\widehat{Y}_{i}^{\test},Y_{i}^{\test})$ with $1\leq i\leq 128$. Similarly, we can also compute those five metrics at the instance-level for each positive bag by using $(\widehat{A}_{im}^{\test}, A_{im})$. Since instance-level labels in negative bags are all negative, those metrics are not computed for negative bags. Then, the average of those metrics can be computed and constitutes our instance-level measures. This leads to a total of $5\times 2=10$ metric-and-level combinations. 

\begin{table}[!h]
\centering
\caption{Prediction results of different methods on the CAMELYON16 dataset. The top and second performances are highlighted in boldface and underline for each column.} 
\label{Tab:pred}
\begin{adjustbox}{max width=\textwidth}
\begin{tabular}{ccccccccccc}
\hline\hline
         & \multicolumn{5}{c}{Bag-level}                                                      & \multicolumn{5}{c}{Instance-level}                                                       \\
         & AUC            & AUPRC          & F1             & Recall         & Precision      & AUC            & AUPRC          & F1             & Recall         & Precision            \\ \hline
SMLE     & \textbf{99.00}          & \textbf{98.43}          & \textbf{94.12}    & \textbf{95.24}    & {\ul 93.02}          & \textbf{98.67}    & \textbf{87.92}    & \textbf{79.93}    & 81.81          & \textbf{83.59} \\
BMLE     & {\ul 98.33}          & 96.14    & {\ul 91.95}          & \textbf{95.24} & 88.89          & {\ul 98.39}          & {\ul 83.75}          & {\ul 76.98}          & {\ul 82.50}          & {\ul 78.00}                \\
ABMIL    & 97.02          & {\ul 97.06}          & 91.57          & 86.36          & \textbf{97.44} & 88.03          & 42.29          & 10.55          & 21.67          & 42.17                \\
TransMIL & 75.82    & 71.10          & 79.84          & 79.84          & 79.84    & 71.55          & 12.92          & 14.51           & 74.33          & 12.70                \\
IRDTV    & 86.32          & 81.59          & 71.70          & 79.17          & 65.52          & 80.08          & 33.09          & 16.08          & 11.32          & 49.33                \\
Its2CLR  & 90.24          & 82.05          & 79.24          & {\ul 86.67}          & 91.30          & 85.56          & 57.65          & 48.87          & \textbf{84.98} & 43.20                \\
PANTHER  & 71.71          & 70.76          & 59.09          & 61.90          & 56.52          & -              & -              & -              & -              & -                    \\ \hline
\end{tabular}
\end{adjustbox}
\end{table}

The detailed results are summarized in Table \ref{Tab:pred}. By Table \ref{Tab:pred}, we find that our estimators (i.e., BMLE and SMLE) both demonstrate excellent predictive performance at both bag and instance levels simultaneously. Among those 10 metric-and-level combinations, SMLE obtains a total of 8 best performances and 1 second-best one. BMLE obtains a total of 7 second-best performances and 1 best one. The overall performances of SMLE and BMLE are clearly better than all their competitors. For example, in terms of bag and instance level AUC, our methods perform clearly better than their competitors. The BMLE achieves the instance-level AUC as high as $98.39\%$, which can be further improved to $98.67\%$ by our SMLE. Considering the potential imbalance in the dataset, we adopt AUPRC and F1 score to further evaluate the methods. Although the ABMIL exhibits competitive performance in bag-level AUPRC, we find that the instance-level AUPRCs of our methods are much better than those of the ABMIL. This illustrates our methods' robustness against potential data imbalance. Since the PANTHER method of \cite{song2024morphological} mainly focuses on the unsupervised slide-level representation, it can hardly be used to detect instance-level labels and therefore we leave it blank in Table \ref{Tab:pred}.
To provide additional insights, we classify the tumor WSIs into two types according to their metastasis sizes (i.e., macro and micro metastasis), and the stratified evaluation results are demonstrated in Appendix B.1.

\begin{figure}[!ht]
\centering 	
\subfigure{\includegraphics[scale=0.35]{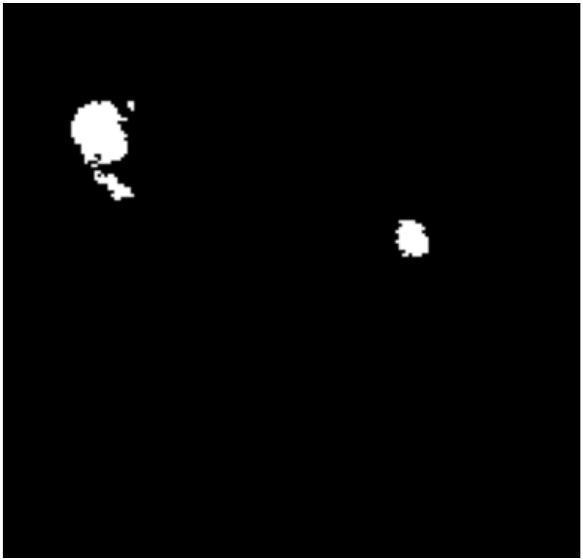}}
\hfil
\subfigure{\includegraphics[scale=0.35]{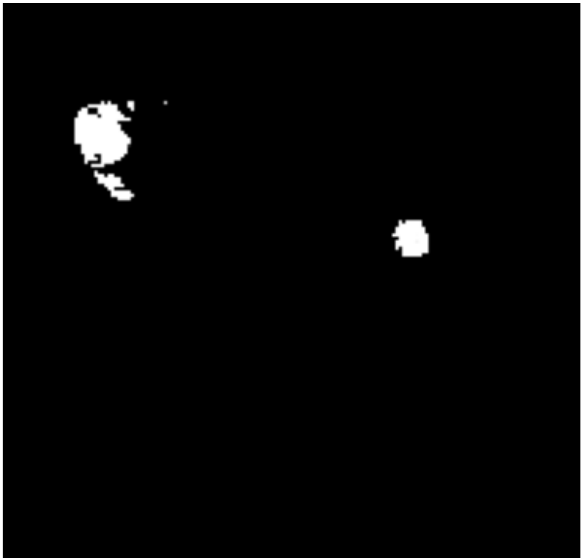}}
\hfil
\subfigure{\includegraphics[scale=0.35]{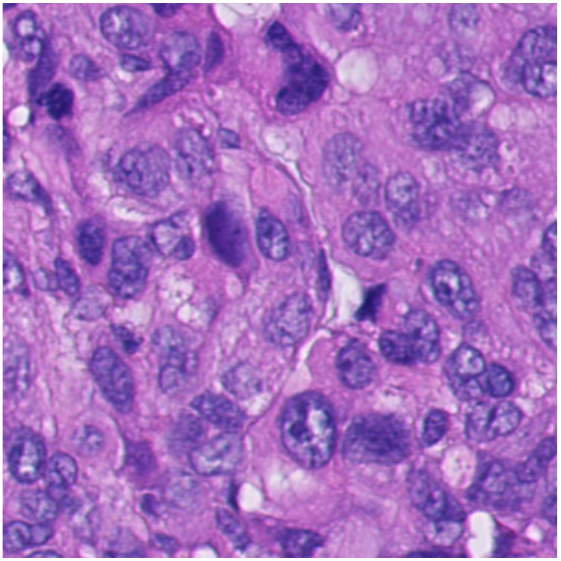}}
\hfil
\subfigure{\includegraphics[scale=0.35]{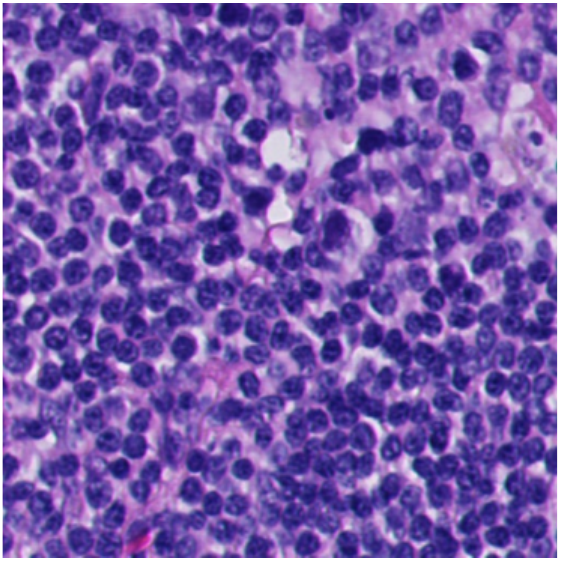}}
\hfil
\caption{The illustration of a test WSI sample. The left panel is the ground truth label of tumors. The middle left panel is the predicted heatmap of tumors. The middle right panel is the sub-image with largest posterior probability of being a tumor. The right panel is the sub-image with the smallest posterior probability of being a tumor.}
\label{Fig:PredImage}
\end{figure}

The left panel of Figure \ref{Fig:PredImage} is the ground truth tumor mask. Each pixel stands for a $512\times 512$ sub-image instance. To gain some intuitive understanding about the instance-level prediction performance, we provide in the middle left panel of Figure \ref{Fig:PredImage} the heatmap of the predicted posterior probability $\widehat{\pi}_{im}^{\test}$ for every sub-image instance.  This should serve as a natural mechanism to identify the significant regions related to the metastasis. In this regard, regions with the largest predicted posterior probability $\widehat{\pi}_{im}^{\test}$ should contribute most to the prediction. By the right two panels of Figure \ref{Fig:PredImage}, we find that sub-images with high predicted probabilities exhibit significant nuclear pleomorphism and a disordered, loosely arranged, and infiltrative growth pattern. This corresponds to regions with classic morphological characteristics of malignancy \citep{rakha2023update}. In contrast, regions with low posterior probabilities align with benign morphological features, such as uniform cell size and shape and a well-circumscribed, dense architecture. 
Therefore, the practitioners can visually interpret clinical needs for explainable AI in pathology in an easier way.

\section{Concluding Remarks}\label{sec:conclude}

To conclude this article, we discuss some intriguing topics for future research. 
First, our theory relies on the GMM assumption for deep theoretical understanding. 
It is of great interest to relax the Gaussian assumption so that the applicability of the theory can be more general. 
Second, given the fact that the percentage of positive instances in a positive bag could be very small, then how to study MIL from the perspective of imbalanced data is another promising direction \citep{zhou2023distributed,li2024gaussian}.
{Third, our current theoretical results only support feature vectors of a fixed dimension. If the covariate dimension is too high (e.g., $q \gg NM$), it seems that overfitting is inevitable. Then, how to modify our model so that features with high dimensions can be allowed is another important topic for future research.}
Lastly, we intend to further integrate the multimodal features by combining the representations from different techniques, thereby bolstering their utility in disease diagnosis. \citep{dai2023orthogonalized}.

\section{Significance Statement}\label{sec:sign}

We study here the problem of WSI analysis for detecting whether breast cancer has spread to lymph nodes. This is of fundamental importance for accurate staging and treatment decisions \citep{ghaznavi2013digital}. An effective method in this regard can help pathologists greatly reduce review time and highlight suspicious tumor regions. However, straightforward analysis is computationally infeasible, since each slide contains billions of pixels. To solve the problem, we divide each slide into many small image patches and develop a unified statistical framework, which can infer labels at both slide and patch levels. The performance of our method can be further improved if a small subset of carefully selected patches is annotated. This makes our method very different from the existing literature.
 
\begin{acks}[Acknowledgments]
The authors would like to thank the anonymous referees, an Associate Editor, and the Editor for their constructive comments that improved the quality of this paper.
\end{acks}

\begin{funding}
Xuetong Li is supported by the China Postdoctoral Science Foundation (No.2025M783133). Jing Zhou is supported by the National Natural Science Foundation of China (No. 72571275) and the Beijing Municipal Social Science Foundation (No. 25BJ03187). Hansheng Wang's research is supported by the National Natural Science Foundation of China (No.12271012 and 72495123).
\end{funding}

\begin{supplement}
\stitle{Supplementary Material of ``Detecting Breast Carcinoma Metastasis on Whole-Slide Images by Partially Subsampled Multiple Instance Learning"}
\sdescription{Appendices A-E in the Supplementary Material provide algorithm development, additional experiments and setups, and proofs as well as technical details of the theorems. The Python module PSMIL can be accessed on PyPI at \url{https://test.pypi.org/project/PSMIL/}. 
The simulation studies and the CAMELYON16 data analysis are available at \url{https://github.com/Jamesyu420/PSMIL}.}
\end{supplement}

\bibliographystyle{imsart-nameyear} 
\bibliography{reference}       

@article{wang2019weakly,
  title={Weakly supervised deep learning for whole slide lung cancer image analysis},
  author={Wang, Xi and Chen, Hao and Gan, Caixia and Lin, Huangjing and Dou, Qi and Tsougenis, Efstratios and Huang, Qitao and Cai, Muyan and Heng, Pheng-Ann},
  journal={IEEE Transactions on Cybernetics},
  volume={50},
  number={9},
  pages={3950--3962},
  year={2019},
  publisher={IEEE}
}

@book{anderson1958introduction,
  title={An Introduction to Multivariate Statistical Analysis},
  author={Anderson, Theodore Wilbur},
  year={2003},
  publisher={Wiley New York}
}

@inproceedings{hou2016patch,
  title={Patch-based convolutional neural network for whole slide tissue image classification},
  author={Hou, Le and Samaras, Dimitris and Kurc, Tahsin M and Gao, Yi and Davis, James E and Saltz, Joel H},
  booktitle={Proceedings of the IEEE Conference on Computer Vision and Pattern Recognition},
  pages={2424--2433},
  year={2016}
}

@inproceedings{li2018thoracic,
  title={Thoracic disease identification and localization with limited supervision},
  author={Li, Zhe and Wang, Chong and Han, Mei and Xue, Yuan and Wei, Wei and Li, Li-Jia and Fei-Fei, Li},
  booktitle={Proceedings of the IEEE Conference on Computer Vision and Pattern Recognition},
  pages={8290--8299},
  year={2018}
}

@article{ghaznavi2013digital,
  title={Digital imaging in pathology: whole-slide imaging and beyond},
  author={Ghaznavi, Farzad and Evans, Andrew and Madabhushi, Anant and Feldman, Michael},
  journal={Annual Review of Pathology: Mechanisms of Disease},
  volume={8},
  pages={331--359},
  year={2013},
  publisher={Annual Reviews}
}

@book{magnus2019matrix,
  title={Matrix Differential Calculus with Applications in Statistics and Econometrics},
  author={Magnus, Jan R and Neudecker, Heinz},
  year={2019},
  publisher={John Wiley \& Sons}
}

@book{van2000asymptotic,
  title={Asymptotic Statistics},
  author={Van der Vaart, Aad W},
  volume={3},
  year={2000},
  publisher={Cambridge University Press}
}

@article{fan2001variable,
  title={Variable selection via nonconcave penalized likelihood and its oracle properties},
  author={Fan, Jianqing and Li, Runze},
  journal={Journal of the American Statistical Association},
  volume={96},
  number={456},
  pages={1348--1360},
  year={2001},
  publisher={Taylor \& Francis}
}

@article{zhou2023ensemble,
  title={An ensemble deep learning model for risk stratification of invasive lung adenocarcinoma using thin-slice CT},
  author={Zhou, Jing and Hu, Bin and Feng, Wei and Zhang, Zhang and Fu, Xiaotong and Shao, Handie and Wang, Hansheng and Jin, Longyu and Ai, Siyuan and Ji, Ying},
  journal={NPJ Digital Medicine},
  volume={6},
  number={1},
  pages={119},
  year={2023},
  publisher={Nature Publishing Group UK London}
}

@article{wang2016deep,
  title={Deep learning for identifying metastatic breast cancer},
  author={Wang, Dayong and Khosla, Aditya and Gargeya, Rishab and Irshad, Humayun and Beck, Andrew H},
  journal={arXiv preprint arXiv:1606.05718},
  year={2016}
}

@article{otsu1979threshold,
  title={A threshold selection method from gray-level histograms},
  author={Otsu, Nobuyuki},
  journal={IEEE Transactions on Systems, Man, and Cybernetics},
  volume={9},
  number={1},
  pages={62--66},
  year={1979},
  publisher={IEEE}
}

@article{bejnordi2017diagnostic,
  title={Diagnostic assessment of deep learning algorithms for detection of lymph node metastases in women with breast cancer},
  author={Bejnordi, Babak Ehteshami and Veta, Mitko and Van Diest, Paul Johannes and Van Ginneken, Bram and Karssemeijer, Nico and Litjens, Geert and Van Der Laak, Jeroen AWM and Hermsen, Meyke and Manson, Quirine F and Balkenhol, Maschenka and others},
  journal={JAMA},
  volume={318},
  number={22},
  pages={2199--2210},
  year={2017},
  publisher={American Medical Association}
}

@article{simonyan2014very,
  title={Very deep convolutional networks for large-scale image recognition},
  author={Simonyan, Karen and Zisserman, Andrew},
  journal={arXiv preprint arXiv:1409.1556},
  year={2014}
}

@inproceedings{stanisavljevic2018fast,
  title={A fast and scalable pipeline for stain normalization of whole-slide images in histopathology},
  author={Stanisavljevic, Milos and Anghel, Andreea and Papandreou, Nikolaos and Andani, Sonali and Pati, Pushpak and Hendrik Ruschoff, Jan and Wild, Peter and Gabrani, Maria and Pozidis, Haralampos},
  booktitle={Proceedings of the European Conference on Computer Vision (ECCV) Workshops},
  year={2018}
}

@article{campanella2019clinical,
  title={Clinical-grade computational pathology using weakly supervised deep learning on whole slide images},
  author={Campanella, Gabriele and Hanna, Matthew G and Geneslaw, Luke and Miraflor, Allen and Werneck Krauss Silva, Vitor and Busam, Klaus J and Brogi, Edi and Reuter, Victor E and Klimstra, David S and Fuchs, Thomas J},
  journal={Nature Medicine},
  volume={25},
  number={8},
  pages={1301--1309},
  year={2019},
  publisher={Nature Publishing Group US New York}
}

@article{wang2022label,
  title={Label cleaning multiple instance learning: Refining coarse annotations on single whole-slide images},
  author={Wang, Zhenzhen and Saoud, Carla and Wangsiricharoen, Sintawat and James, Aaron W and Popel, Aleksander S and Sulam, Jeremias},
  journal={IEEE Transactions on Medical Imaging},
  volume={41},
  number={12},
  pages={3952--3968},
  year={2022},
  publisher={IEEE}
}

@article{liu2021statistical,
  title={Statistical disease mapping for heterogeneous neuroimaging studies},
  author={Liu, Rongjie and Zhu, Hongtu and Alzheimer's Disease Neuroimaging Initiative},
  journal={Canadian Journal of Statistics},
  volume={49},
  number={1},
  pages={10--34},
  year={2021},
  publisher={Wiley Online Library}
}

@inproceedings{zhu2017wsisa,
  title={Wsisa: Making survival prediction from whole slide histopathological images},
  author={Zhu, Xinliang and Yao, Jiawen and Zhu, Feiyun and Huang, Junzhou},
  booktitle={Proceedings of the IEEE Conference on Computer Vision and Pattern Recognition},
  pages={7234--7242},
  year={2017}
}

@book{diggle2013statistical,
  title={Statistical Analysis of Spatial and Spatio-temporal Point Patterns},
  author={Diggle, Peter J},
  year={2013},
  publisher={CRC press}
}

@article{waks2019breast,
  title={Breast cancer treatment: A review},
  author={Waks, Adrienne G and Winer, Eric P},
  journal={JAMA},
  volume={321},
  number={3},
  pages={288--300},
  year={2019},
  publisher={American Medical Association}
}

@inproceedings{liu2012key,
  title={Key instance detection in multi-instance learning},
  author={Liu, Guoqing and Wu, Jianxin and Zhou, Zhi-Hua},
  booktitle={Asian Conference on Machine Learning},
  pages={253--268},
  year={2012},
  organization={PMLR}
}

@article{Bray2024GlobalCS,
  title={Global cancer statistics 2022: GLOBOCAN estimates of incidence and mortality worldwide for 36 cancers in 185 countries},
  author={Freddie Bray and Mathieu Laversanne and Hyuna Sung and Jacques Ferlay and Rebecca L. Siegel and Isabelle Soerjomataram and Ahmedin Jemal},
  journal={CA: A Cancer Journal for Clinicians},
  year={2024},
  volume={74},
  pages={229 - 263},
  url={https://api.semanticscholar.org/CorpusID:268886842}
}

@article{withers2021x,
  title={X-ray computed tomography},
  author={Withers, Philip J and Bouman, Charles and Carmignato, Simone and Cnudde, Veerle and Grimaldi, David and Hagen, Charlotte K and Maire, Eric and Manley, Marena and Du Plessis, Anton and Stock, Stuart R},
  journal={Nature Reviews Methods Primers},
  volume={1},
  number={1},
  pages={18},
  year={2021},
  publisher={Nature Publishing Group UK London}
}

@article{chen2024towards,
title={Towards a general-purpose foundation model for computational pathology},
author={Chen, Richard J and Ding, Tong and Lu, Ming Y and Williamson, Drew FK and Jaume, Guillaume and Song, Andrew H and Chen, Bowen and Zhang, Andrew and Shao, Daniel and Shaban, Muhammad and others},
journal={Nature Medicine},
volume={30},
number={3},
pages={850--862},
year={2024},
publisher={Nature Publishing Group US New York}
}

@article{li2024gaussian,
  author={Li, Xuetong and Zhou, Jing and Wang, Hansheng},
  title   = {Gaussian mixture models with rare events},
  journal = {Journal of Machine Learning Research},
  year    = {2024},
  volume  = {25},
  number  = {252},
  pages   = {1--40},
  url     = {http://jmlr.org/papers/v25/23-1245.html}
}

@inproceedings{ilse2018attention,
  title={Attention-based deep multiple instance learning},
  author={Ilse, Maximilian and Tomczak, Jakub and Welling, Max},
  booktitle={International Conference on Machine Learning},
  pages={2127--2136},
  year={2018},
  organization={PMLR}
}

@book{rangayyan2024biomedical,
  title={Biomedical Signal Analysis},
  author={Rangayyan, Rangaraj M and Krishnan, Sridhar},
  year={2024},
  publisher={John Wiley \& Sons}
}

@inproceedings{liu2023multiple,
  title={Multiple instance learning via iterative self-paced supervised contrastive learning},
  author={Liu, Kangning and Zhu, Weicheng and Shen, Yiqiu and Liu, Sheng and Razavian, Narges and Geras, Krzysztof J and Fernandez-Granda, Carlos},
  booktitle={Proceedings of the IEEE/CVF Conference on Computer Vision and Pattern Recognition},
  pages={3355--3365},
  year={2023}
}

@article{krag1998sentinel,
  title={The sentinel node in breast cancer-a multicenter validation study},
  author={Krag, David and Weaver, Donald and Ashikaga, Takamaru and Moffat, Frederick and Klimberg, V Suzanne and Shriver, Craig and Feldman, Sheldon and Kusminsky, Roberto and Gadd, Michele and Kuhn, Joseph and others},
  journal={New England Journal of Medicine},
  volume={339},
  number={14},
  pages={941--946},
  year={1998},
  publisher={Mass Medical Soc}
}

@article{ibrahim2020artificial,
  title={Artificial intelligence in digital breast pathology: techniques and applications},
  author={Ibrahim, Asmaa and Gamble, Paul and Jaroensri, Ronnachai and Abdelsamea, Mohammed M and Mermel, Craig H and Chen, Po-Hsuan Cameron and Rakha, Emad A},
  journal={The Breast},
  volume={49},
  pages={267--273},
  year={2020},
  publisher={Elsevier}
}

@article{elmore2015diagnostic,
  title={Diagnostic concordance among pathologists interpreting breast biopsy specimens},
  author={Elmore, Joann G and Longton, Gary M and Carney, Patricia A and Geller, Berta M and Onega, Tracy and Tosteson, Anna NA and Nelson, Heidi D and Pepe, Margaret S and Allison, Kimberly H and Schnitt, Stuart J and others},
  journal={JAMA},
  volume={313},
  number={11},
  pages={1122--1132},
  year={2015},
  publisher={American Medical Association}
}

@article{shao2021transmil,
  title={Transmil: Transformer based correlated multiple instance learning for whole slide image classification},
  author={Shao, Zhuchen and Bian, Hao and Chen, Yang and Wang, Yifeng and Zhang, Jian and Ji, Xiangyang and others},
  journal={Advances in Neural Information Processing Systems},
  volume={34},
  pages={2136--2147},
  year={2021}
}

@inproceedings{he2016deep,
  title={Deep residual learning for image recognition},
  author={He, Kaiming and Zhang, Xiangyu and Ren, Shaoqing and Sun, Jian},
  booktitle={Proceedings of the IEEE Conference on Computer Vision and Pattern Recognition},
  pages={770--778},
  year={2016}
}

@inproceedings{zhang2018shufflenet,
  title={ShuffleNet: An extremely efficient convolutional neural network for mobile devices},
  author={Zhang, Xiangyu and Zhou, Xinyu and Lin, Mengxiao and Sun, Jian},
  booktitle={Proceedings of the IEEE Conference on Computer Vision and Pattern Recognition},
  pages={6848--6856},
  year={2018}
}

@inproceedings{dosovitskiy2020image,
  title={An image is worth 16x16 words: Transformers for image recognition at scale},
  author={Dosovitskiy, Alexey and Beyer, Lucas and Kolesnikov, Alexander and Weissenborn, Dirk and Zhai, Xiaohua and Unterthiner, Thomas and Dehghani, Mostafa and Minderer, Matthias and Heigold, Georg and Gelly, Sylvain and others},
  booktitle={International Conference on Learning Representations},
  year={2020}
}

@article{oquab2023dinov2,
  title={Dinov2: Learning robust visual features without supervision},
  author={Oquab, Maxime and Darcet, Timoth{\'e}e and Moutakanni, Th{\'e}o and Vo, Huy and Szafraniec, Marc and Khalidov, Vasil and Fernandez, Pierre and Haziza, Daniel and Massa, Francisco and El-Nouby, Alaaeldin and others},
  journal={arXiv preprint arXiv:2304.07193},
  year={2023}
}

@article{carbonneau2018multiple,
  title={Multiple instance learning: A survey of problem characteristics and applications},
  author={Carbonneau, Marc-Andr{\'e} and Cheplygina, Veronika and Granger, Eric and Gagnon, Ghyslain},
  journal={Pattern Recognition},
  volume={77},
  pages={329--353},
  year={2018},
  publisher={Elsevier}
}

@article{lee2022derivation,
  title={Derivation of prognostic contextual histopathological features from whole-slide images of tumours via graph deep learning},
  author={Lee, Yongju and Park, Jeong Hwan and Oh, Sohee and Shin, Kyoungseob and Sun, Jiyu and Jung, Minsun and Lee, Cheol and Kim, Hyojin and Chung, Jin-Haeng and Moon, Kyung Chul and others},
  journal={Nature Biomedical Engineering},
  pages={1--15},
  year={2022},
  publisher={Nature Publishing Group UK London}
}

@article{zhang2023image,
  title={Image response regression via deep neural networks},
  author={Zhang, Daiwei and Li, Lexin and Sripada, Chandra and Kang, Jian},
  journal={Journal of the Royal Statistical Society Series B: Statistical Methodology},
  volume={85},
  number={5},
  pages={1589--1614},
  year={2023},
  publisher={Oxford University Press US}
}

@article{dai2023orthogonalized,
  title={Orthogonalized kernel debiased machine learning for multimodal data analysis},
  author={Dai, Xiaowu and Li, Lexin},
  journal={Journal of the American Statistical Association},
  volume={118},
  number={543},
  pages={1796--1810},
  year={2023},
  publisher={Taylor \& Francis}
}

@article{liu2024robust,
  title={Robust high-dimensional regression with coefficient thresholding and its application to imaging data analysis},
  author={Liu, Bingyuan and Zhang, Qi and Xue, Lingzhou and Song, Peter X-K and Kang, Jian},
  journal={Journal of the American Statistical Association},
  volume={119},
  number={545},
  pages={715--729},
  year={2024},
  publisher={Taylor \& Francis}
}

@article{zhou2023distributed,
  title={Distributed empirical likelihood approach to integrating unbalanced datasets},
  author={Zhou, Ling and She, Xichen and Song, Peter X-K},
  journal={Statistica Sinica},
  volume={33},
  number={3},
  pages={2209--2231},
  year={2023},
  publisher={JSTOR}
}

@article{deng2024optimal,
  title={Optimal and safe estimation for high-dimensional semi-supervised learning},
  author={Deng, Siyi and Ning, Yang and Zhao, Jiwei and Zhang, Heping},
  journal={Journal of the American Statistical Association},
  volume={119},
  number={548},
  pages={2748--2759},
  year={2024},
  publisher={Taylor \& Francis}
}

@article{patz2014overdiagnosis,
  title={Overdiagnosis in low-dose computed tomography screening for lung cancer},
  author={Patz, Edward F and Pinsky, Paul and Gatsonis, Constantine and Sicks, JoRean D and Kramer, Barnett S and Tammem{\"a}gi, Martin C and Chiles, Caroline and Black, William C and Aberle, Denise R and NLST Overdiagnosis Manuscript Writing Team and others},
  journal={JAMA Internal Medicine},
  volume={174},
  number={2},
  pages={269--274},
  year={2014},
  publisher={American Medical Association}
}

@article{mohammed2021radiohead,
  title={RADIOHEAD: Radiogenomic analysis incorporating tumor heterogeneity in imaging through densities},
  author={Mohammed, Shariq and Bharath, Karthik and Kurtek, Sebastian and Rao, Arvind and Baladandayuthapani, Veerabhadran},
  journal={The Annals of Applied Statistics},
  volume={15},
  number={4},
  pages={1808--1830},
  year={2021},
  publisher={JSTOR}
}

@article{houssein2021deep,
  title={Deep and machine learning techniques for medical imaging-based breast cancer: A comprehensive review},
  author={Houssein, Essam H and Emam, Marwa M and Ali, Abdelmgeid A and Suganthan, Ponnuthurai Nagaratnam},
  journal={Expert Systems with Applications},
  volume={167},
  pages={114161},
  year={2021},
  publisher={Elsevier}
}

@inproceedings{song2024morphological,
  title={Morphological prototyping for unsupervised slide representation learning in computational pathology},
  author={Song, Andrew H and Chen, Richard J and Ding, Tong and Williamson, Drew FK and Jaume, Guillaume and Mahmood, Faisal},
  booktitle={Proceedings of the IEEE/CVF Conference on Computer Vision and Pattern Recognition},
  pages={11566--11578},
  year={2024}
}

@article{wu2024individualized,
  title={Individualized image region detection with total variation},
  author={Wu, Sanyou and Wang, Fuying and Feng, Long},
  journal={Statistical Analysis and Data Mining: The ASA Data Science Journal},
  volume={17},
  number={3},
  pages={e11684},
  year={2024},
  publisher={Wiley Online Library}
}

@article{kawada2011significance,
  title={Significance and mechanism of lymph node metastasis in cancer progression},
  author={Kawada, Kenji and Taketo, Makoto M},
  journal={Cancer Research},
  volume={71},
  number={4},
  pages={1214--1218},
  year={2011},
  publisher={American Association for Cancer Research}
}

@article{cucinella2024prognostic,
  title={Prognostic value of isolated tumor cells in sentinel lymph nodes in low risk endometrial cancer: results from an international multi-institutional study},
  author={Cucinella, Giuseppe and Schivardi, Gabriella and Zhou, Xun Clare and AlHilli, Mariam and Wallace, Sumer and Wohlmuth, Christoph and Baiocchi, Glauco and Tokgozoglu, Nedim and Raspagliesi, Francesco and Buda, Alessandro and others},
  journal={International Journal of Gynecological Cancer},
  volume={34},
  number={2},
  pages={179--187},
  year={2024},
  publisher={Elsevier}
}

@article{rakha2023update,
  title={An update on the pathological classification of breast cancer},
  author={Rakha, Emad A and Tse, Gary M and Quinn, Cecily M},
  journal={Histopathology},
  volume={82},
  number={1},
  pages={5--16},
  year={2023},
  publisher={Wiley Online Library}
}

@article{yu2025semiparametric,
  title={A semiparametric Gaussian Mixture Model with spatial dependence and its application to whole-slide image clustering analysis},
  author={Yu, Baichen and Liu, Jin and Wang, Hansheng},
  journal={Biometrics},
  volume={81},
  number={4},
  pages={ujaf149},
  year={2025},
  publisher={Oxford University Press}
}

@article{rubin1976inference,
  title={Inference and missing data},
  author={Rubin, Donald B},
  journal={Biometrika},
  volume={63},
  number={3},
  pages={581--592},
  year={1976},
  publisher={Oxford University Press}
}

@book{little2019statistical,
  title={Statistical Analysis with Missing Data},
  author={Little, Roderick JA and Rubin, Donald B},
  year={2019},
  publisher={John Wiley \& Sons}
}

@article{besag1986statistical,
  title={On the statistical analysis of dirty pictures},
  author={Besag, Julian},
  journal={Journal of the Royal Statistical Society Series B: Statistical Methodology},
  volume={48},
  number={3},
  pages={259--279},
  year={1986},
  publisher={Oxford University Press}
}

\end{document}